\documentclass[]{aastex631}

\usepackage{graphicx}
\usepackage{subfigure}

\begin{document}

\title{Probing the cosmological 21~cm global signal from the Antarctic ice sheet}

\author[0000-0003-2775-3523]{Shijie Sun}
\affiliation{National Astronomical Observatories, Chinese Academy of Sciences, Beijing 100101, China}

\author{Jiaqin Xu}
\affiliation{National Astronomical Observatories, Chinese Academy of Sciences, Beijing 100101, China}

\author{Minquan Zhou}
\affiliation{School of Mechanical Engineering, Hangzhou Dianzi University, Hangzhou 310018, China}

\author{Shenzhe Xu}
\affiliation{School of Mechanical Engineering, Hangzhou Dianzi University, Hangzhou 310018, China}

\author{Fengquan Wu}
\affiliation{National Astronomical Observatories, Chinese Academy of Sciences, Beijing 100101, China}

\author{Haoran Zhang}
\affiliation{College of Physics and Electronic Engineering, Shanxi University, Taiyuan 030006, China}

\author{Juyong Zhang}
\affiliation{School of Mechanical Engineering, Hangzhou Dianzi University, Hangzhou 310018, China}

\author{Bin Ma}
\affiliation{School of Physics and Astronomy, Sun Yat-sen University, Zhuhai 519082, China}

\author{Zhaohui Shang}
\affiliation{National Astronomical Observatories, Chinese Academy of Sciences, Beijing 100101, China}

\author{Xuelei Chen}
\correspondingauthor{Xuelei Chen}
\affiliation{National Astronomical Observatories, Chinese Academy of Sciences, Beijing 100101, China}
\affiliation{School of Astronomy and Space Science, University of Chinese Academy of Sciences, Beijing 100049, China}
\affiliation{Center of High Energy Physics, Peking University, Beijing 100871, China}
\email{xuelei@cosmology.bao.ac.cn}



\begin{abstract}

The redshifted 21~cm line, arising from neutral hydrogen, offers a unique probe into the intergalactic medium and the first stars and galaxies formed in the early universe. However, detecting this signal is a challenging task because of artificial radio-frequency interference (RFI) and systematic errors such as ground effects. The interior of the Antarctic continent provides an excellent location to make such observations, with minimal RFI and relatively stable foreground signals. Moreover, a flat plateau in central Antarctica, with an ice cap over 2000 m deep, will show less ground reflection of radio waves, reducing the signal complexity in the area around the probing antenna.
 It may be advantageous to perform cosmological 21~cm experiments in Antarctica, and a 21~cm Antarctic global spectrum experiment can potentially be deployed on the Antarctic ice cap. We have performed preliminary instrumental design, system calibration, and implementation of such an instrument optimized for extreme cold and capable of long-term autonomous operation. This system shows the ability to effectively detect the 21~cm signal, confirming Antarctica as an excellent observational site for radio cosmology. 

\end{abstract}



\section{Introduction} 
\label{sec:intro}

The redshifted 21~cm line, arising from neutral hydrogen, is potentially observable at low radio frequencies (50~--~200~MHz) and carries much information about the evolutionary history of the cosmic dawn and the epoch of reionization~\citep{madau199721,Chen2004,Chen_2008,furlanetto2006cosmology,barkana2016rise}.
The global (all-sky averaged) spectrum of the redshifted 21~cm brightness temperature can be measured with a single antenna and a wide-band spectrometer, and numerous experimental efforts are devoted to this, including the Experiment to Detect the Global Epoch of reionization Signature (EDGES,~\citet{bowman2008toward,Bowman2010}), the Broadband Instrument for Global HydrOgen ReioNization Signal (BIGHORNS,~\citet{sokolowski2015bighorns}), the Shaped Antenna measurement of the background RAdio Spectrum (SARAS,~\citet{patra2013saras,saras2018,saras3}), the Probing Radio Intensity at high-Z from Marion (${\rm PRI^ZM}$,~\citet{philip2019probing}), Radio Experiment for the Analysis of Cosmic Hydrogen (REACH,~\citet{de2019reach}), the Large-aperture Experiment to Detect the Dark Age (LEDA,~\citet{price2018design}), Sonda Cosmologica de las Islas para la Deteccion de Hidrogeno Neutro (SCI-HI,~\citet{Voytek2014}), Cosmic Twilight Polarimeter (CTP,~\citet{Nhan2019}), and Mapper of the IGM Spin Temperature (MIST,~\citet{MIST-web}). There are also experiments using short-spaced interferometers, such as the Short-Spaced Interferometer Telescope probing cosmic dAwn and epoch of ReionizAtion (SITARA, ~\citet{Thekkeppattu_McKinley_Trott_Jones_Ung_2022}).
Considering the measurement difficulty from most ground sites at low frequencies, caused by ionospheric refraction and reflection of broadband radio-frequency interference (RFI), there are also proposals to make the measurement in the lunar orbit, such as the Dark Ages Radio Explorer project (DARE,~\citet{burns2017space}), the Dark Ages Polarimetry PathfindER in low lunar orbit (DAPPER,~\citet{burns2021global}), and the Discovering the Sky at Longest wavelength project (DSL,~\cite{Chen:2020lok,chen2023,Shi2022}). 

Depending on the cosmological model, the
depth of the global cosmic dawn signal is $\sim$100~mK. Measuring this faint signal is extremely challenging, because the galactic foreground, which dominates the radio sky, is more than four orders of magnitude brighter than the signal to be measured. Any small system effect can either swamp the signal or produce a false signal. Extremely high sensitivity and large dynamic ranges are required to discern the small 21~cm signature in the spectrum, and the 21~cm signal from the cosmic dawn has an unknown but generally broad shape. Its detection requires a good understanding of the instrument response, which must be determined by a calibration procedure. Any frequency dependence in instrumental gain, noise spectrum, antenna beam shape, ionospheric effect, and ground reflection or other effects, if not properly accounted for, can affect the measurement result. 

The EDGES experiment has reported the detection of a 500~mK deep absorption feature centered at 78~MHz ~\citep{bowman2018absorption}, which may be associated with the signature of the cosmic dawn. This absorption is much stronger than the prediction made by the standard model,  so its cosmological interpretation may require new physics or astrophysics mechanisms, e.g., a cooling mechanism with exotic dark matter particles~\citep{barkana2018possible} or extra radio background~\citep{yang2020abundance,berlin2018severely}. However, it has also been questioned whether this feature is real or arises from systematic errors~\citep{hills2018concerns, bradley2019ground, sun2024calibration}. Measurements taken by the SARAS-3 experiment have not detected such an absorption feature~\citep{saras2021}.

The ground effect is one of the most significant systematic factors in a ground-based experiment. The proximity of the antenna to the ground surface causes antenna-ground coupling, which affects performance by altering the impedance, radiation pattern, and frequency response of the antenna. In addition, below the antenna, the ground generally has multiple layers of soil or rock, and can reflect incoming radio waves at the interfaces of these layers. A reflected wave can interfere with the direct signals from the sky, forming complicated standing waves, and causing a multi-path effect which distorts the data. The ground also emits thermal radiation, which can add noise to the measurements. All of these effects also depend on the temperature of the ground, which changes over time.  

Experiments have taken various measures to address these issues, such as precise modeling of the ground in the data analysis or simplifying the antenna--ground coupling by applying a large conducting ground screen under the antenna~\citep{bowman2018absorption}. Taking advantage of any special condition inherent to the terrain is also a route worth exploring. For example, the SARAS experiment team placed their antenna above a water surface~\citep{saras2021}, and it may also be possible to hang the measurement instrument in a deep canyon.  

The continental interior of Antarctica may offer an ideal site for low-frequency radio astronomy requiring extreme detection accuracy, such as a 21~cm global spectrum experiment, on account of its remote, dry, and stable environment, with the Antarctic ice helping to reduce ground reflections. Here, we investigate this possibility. 
In Section~\ref{sec:environment} we describe the advantages of the Antarctic inland for low-frequency radio astronomy, with an emphasis on global spectrum experiments. In Section~\ref{sec:GSM} we present the design of a global spectrum measurement instrument for use in Antarctica, including the antenna, receiver, calibration mechanism, receiver case, and power supply. In Section~\ref{sec:results} we present the electromagnetic environment and ground-penetrating radar measurement results in inland Antarctica, and describe the implementation of our instrument. Section~\ref{sec:conclusions} gives our conclusions and discusses future work.

\section{Antarctica for astronomy} 
\label{sec:environment}

\subsection{The advantages of Antarctica for astronomical observations}
Antarctica, especially its inland plateaus (Dome A, Dome C, and the South Pole) has long been recognized as offering a number of unique advantages for astronomical observations~\citep{Shang2020}. For example, the air over inland Antarctica is cold, stable, and very dry, and at the high elevations of the plateaus the air is also thinner; the distortion, absorption and scattering of light by the air are all significantly reduced. The absence of water vapor is particularly beneficial for infrared and submillimeter wavelengths.  The long polar nights which last for months allow for long, uninterrupted darkness, which is ideal for time domain astronomy projects, which require continuous monitoring of celestial objects. During night Antarctica is one of the darkest places on Earth, with effectively no artificial light to interfere with observations. A number of astronomical observatories have been set up there,  such as the South Pole Telescope \citep{ruhl2004south, carlstrom201110}, the Background Imaging of Cosmic Extragalactic Polarization (BICEP) experiment for studying the cosmic microwave background~\citep{keating2003bicep}, the Antarctic Search for Transiting ExoPlanets (ASTEP, \citet{crouzet2010astep}), a 40~cm optical telescope at Dome C, and the Antarctic Survey Telescope (AST3), which is a series of optical telescopes designed for wide-field sky surveys, at Dome A \citep{yuan2012optical, yuan2014ast3}, which has the highest elevation in Antarctica. This has been investigated as one of the best sites on Earth for optical observations~\citep{ma2020night, yang2021cloud}, as well as for far-infrared observations~\citep{shi2016terahertz}.

\begin{figure}[htbp]
\centering
\includegraphics[width=0.5\columnwidth]{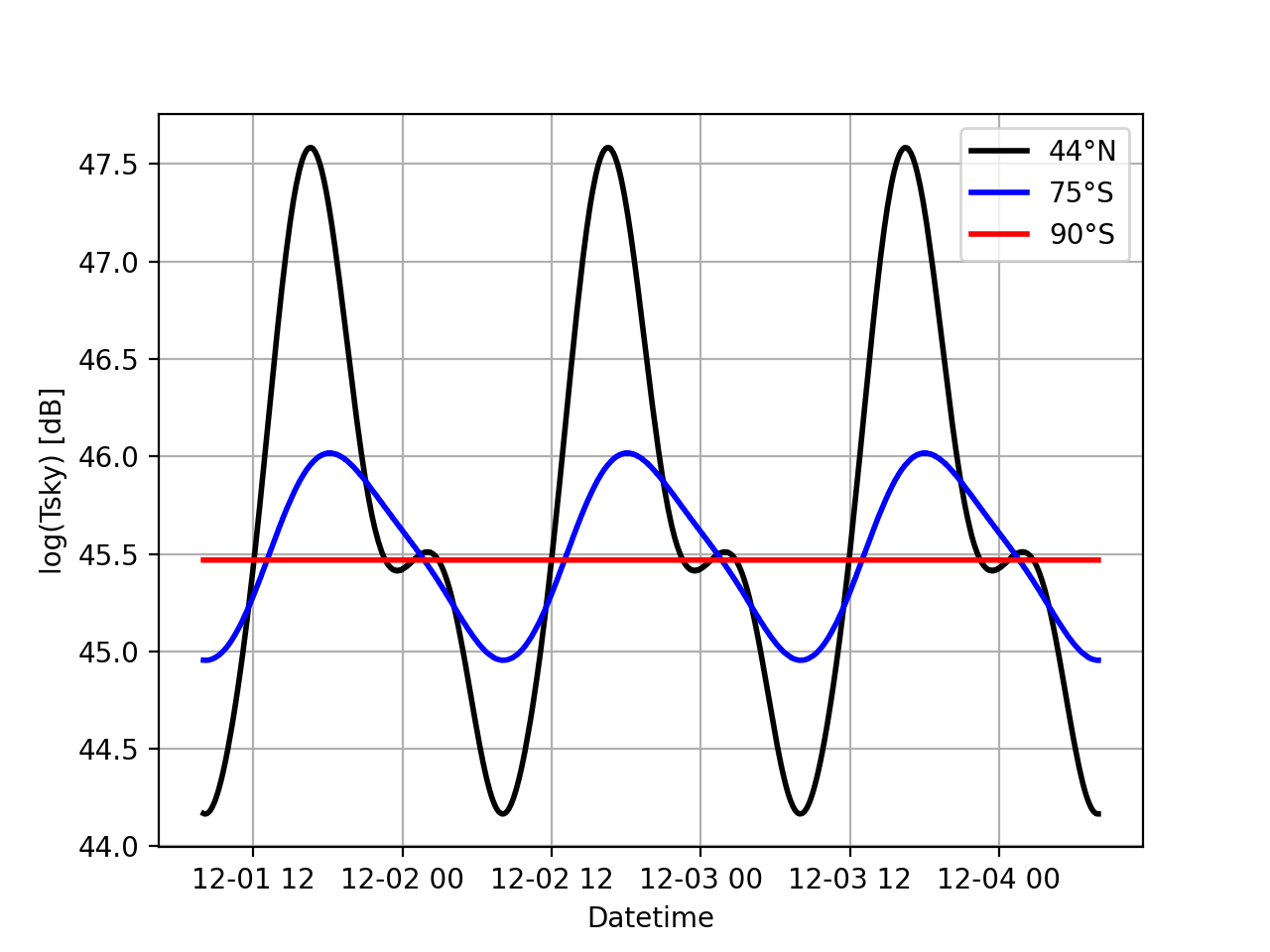}
\caption{The beam-average sky temperature as a function of time for several latitudes.} 
\label{fig:sky temp vs latitude}
\end{figure}

Here, we investigate the potential advantages of the Antarctic inland plateaus for low-frequency radio astronomy. First, in the low-frequency band, RFI is generally severe for astronomical observations, because a variety of human technological activities generate low-frequency emissions. The remote and sparsely populated nature of Antarctica, where such activities are minimal or nonexistent, consequently makes it an ideal location for observing faint cosmic signals. In addition, the inland of Antarctica is covered with a thick ice sheet, with a depth of approximately 2000 meters. This may provide relatively stable and uniform ground for the 21~cm global spectrum experiment, allowing antenna--ground coupling and standing wave effects can be reduced. Finally, toward the geographic pole, most sidereal motion is apparent as rotation of the sky, with fewer celestial objects rising above or setting below the horizon, thus providing a more stable sky for global spectrum measurements. These potential benefits may make the Antarctic inland plateaus one of the best terrestrial sites for 21~cm global spectrum measurement.

Here, we look into the stable sky signal (the RFI environment and the ice sheet ground will be discussed in further detail in Section~\ref{sec:GSM}). Antennas used for measuring radio signals have some chromatic effect, because the antenna beam profile changes with frequency. The anisotropy of the sky foreground can cause spectral variations that affect global spectrum measurements and need to be corrected according to the sky model and antenna beam, which is a major source of error in the cosmic dawn 21~cm experiment. One can attempt to minimize the chromatic effect through antenna design, but in practice it is almost impossible to completely eliminate. In the vicinity of the South Pole, the South Celestial Pole is close to the zenith, so that most of the visible sky area is located within the circle of perpetual apparition, and there is no obvious ascent or descent of most celestial bodies. When an antenna of wide beam is used, especially if the center of the field of view is pointing at the zenith, there is little change in antenna temperature with time. Fig.~\ref{fig:sky temp vs latitude} shows the beam-averaged sky signal as a function of time at several latitudes. At 44\textdegree north (e.g., a site in Xinjiang, China), the temperature variation curve (black line) is complicated, and the reduction of such a foreground would require an accurate model of the sky and a highly precise antenna beam, which is difficult to achieve in practice. At 75\textdegree south (blue line), the variation of the beam-averaged temperature is much simpler and its amplitude is only about 1~dB. At the South Pole, there is no variation in the signal, as apparent motion is solely rotation for all objects except the Sun and planets, with targets never rising or setting. As a result, provided that the system has cylindrical symmetry around its vertical axis, it is almost immune to chromatic error. 

\subsection{Expedition, site selection and implementation}
\label{subsec:implementation}


Despite the many advantages of the Antarctic sites, there are many challenges when doing astronomy in Antarctica. Notably, it is extremely cold in Antarctica, with temperatures falling below –80 °C, which makes equipment maintenance and human operation difficult, and puts high requirements on instrument reliability. Antarctic sites are also very remote and isolated, meaning that instruments must rely heavily on automated systems, and have a self-sufficient energy supply. Finally, and perhaps most importantly, inland Antarctica has extremely limited physical accessibility, and it is not always possible to transport people and equipment to such sites.  

\begin{figure}[htbp]
\centering
\includegraphics[width=0.5\columnwidth]{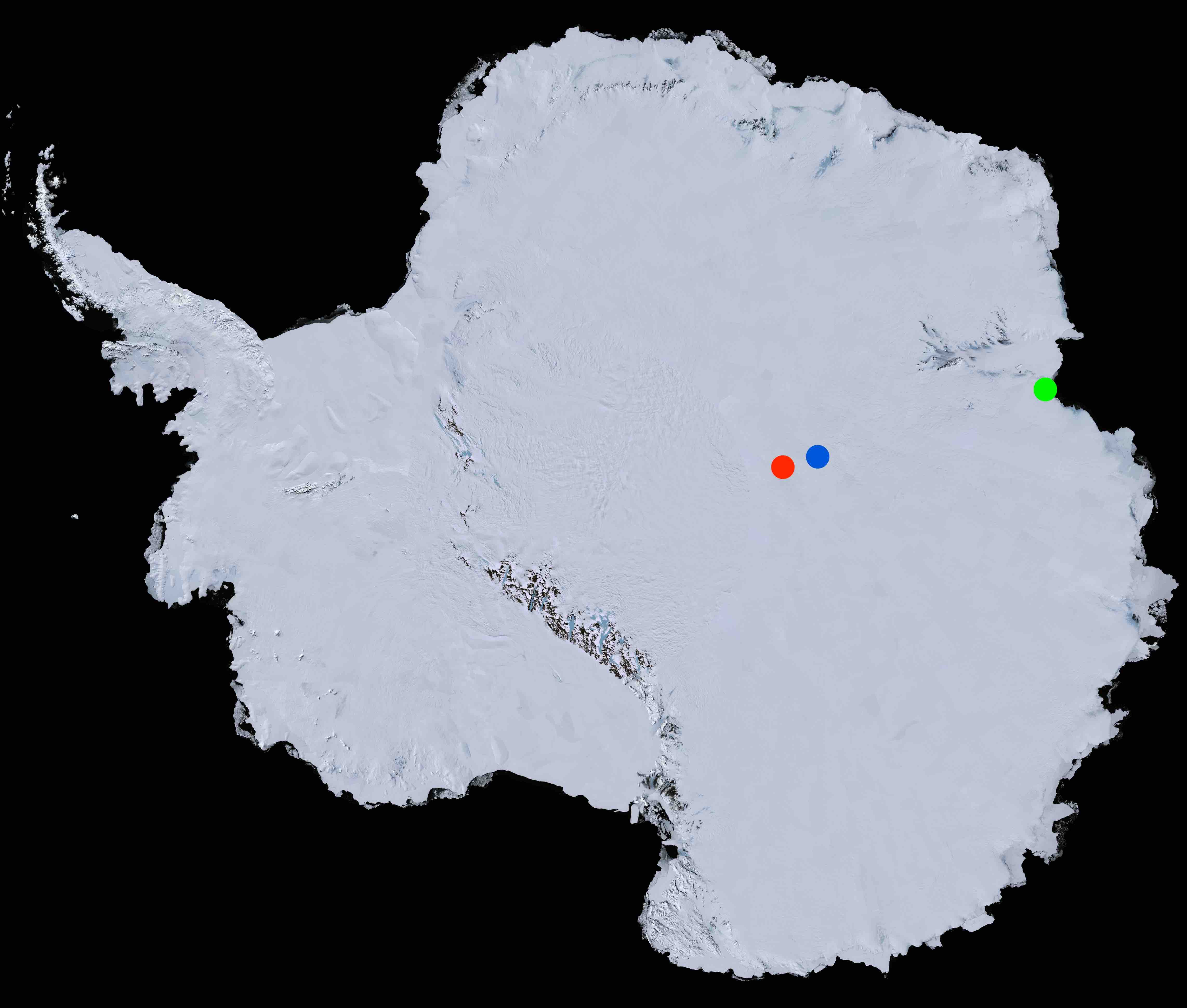}
\caption{A map of Antarctica, showing the position of Zhongshan Station (green dot), Kunlun Station (red dot), and the site of our 21~cm global spectrum experiment (blue dot) at (78°32'11"S, 77°0'50"E).}
\label{fig:position}
\end{figure}

The Chinese National Antarctica Research Expedition program offers a precious opportunity to conduct a low-frequency radio astronomical experiment in inland Antarctica. Several scientific research stations have been set up on the Antarctic continent to provide facilities for scientific research. Kunlun Station, first established in 2009, is located in Dome A (80°25'01"S, 77°06'58"E), the highest plateau in Antarctica (see Fig.~\ref{fig:position}). Many scientific instruments for astronomy and other scientific disciplines are installed at the site, and an inland expedition visits this site each year, during the southern hemisphere summer, to install new instruments and maintain or upgrade the old ones. Equipment for a 21~cm global spectrum experiment on the Antarctic inland ice cap can be installed during such an expedition. 

Our site selection is guided by three primary principles: the chosen location should be situated on the Antarctic inland ice cap, with an ice thickness exceeding 100 meters; the site should be characterized by flat and open terrain; and it should not be in close proximity to pre-existing installations on the Antarctic ice cap, such as meteorological stations and geomagnetic instruments, as these may introduce electromagnetic interference.

In December 2023, the 40th Chinese Antarctic research team inland expedition departed from Zhongshan Station (69°22'24.76"S, 76°22'14.28"E) toward Kunlun Station (80°25'01"S, 77°06'58"E), advancing approximately 100 km per day. The research team conducted site selection for our global spectrum experiment along the route, measuring RFI at flat potential sites, and surveying the ice layers using ground-penetrating radar (GPR). We have selected a site at (78°32'11"S, 77°0'50"E), approximately 1050 km from Zhongshan Station, for the installation of the Antarctic global spectrum observation instrument (see Fig.~\ref{fig:position}). 

\begin{figure}[htbp]
\centering
\includegraphics[width=0.5\columnwidth]{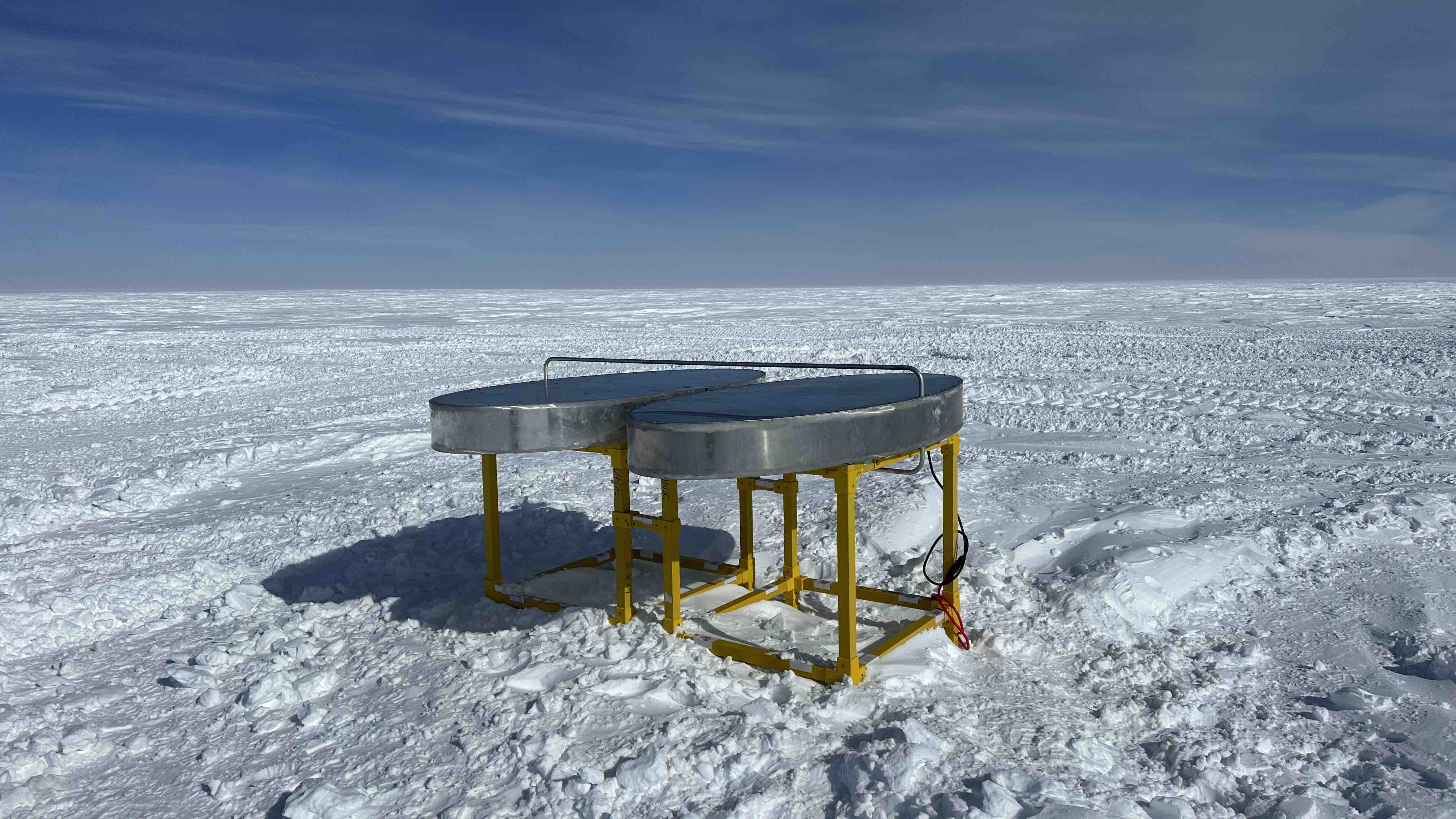}
\caption{The Antarctic global spectrum measurement instrument.}
\label{fig:in_situ_instrument}
\end{figure}
\begin{figure}[htbp]
\centering
\includegraphics[width=0.5\columnwidth]{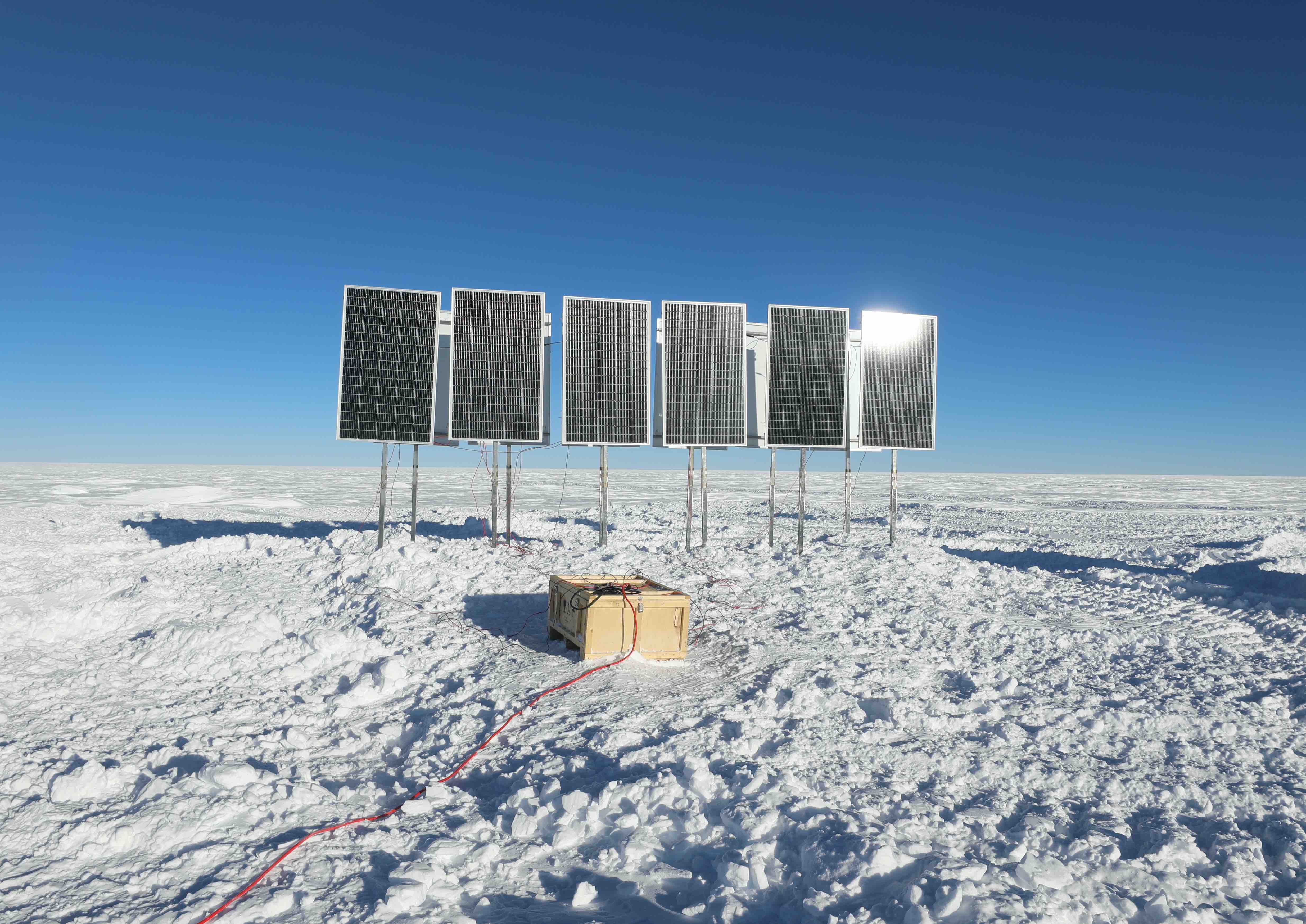}
\caption{The solar Panels.} 
\label{fig:in_situ_solarpanel}
\end{figure}

At the selected site, the expedition team excavated a pit approximately 1~m deep with a level base. They then embedded the supporting bracket of the antenna of the global spectrum experiment into the pit and buried the bracket with ice after the antenna and the receiver were installed. The installed antenna is shown in Fig.~\ref{fig:in_situ_instrument}.

To supply power for the operation of the instrument, we installed a set of solar panels, which are placed approximately 500~m from the antenna itself, to avoid affecting any measurements. There are 12 solar panels, with sets of three connected in series to provide 100~V each, making four groups connected by shunt. Given the prevalence of strong winds on the Antarctic ice cap, these solar panels were strategically aligned parallel to the wind direction to mitigate snow accumulation. The configuration of the solar panel array is shown in Fig.~\ref{fig:in_situ_solarpanel}. 

We selected spiral lead-acid batteries for energy storage, with a relatively wide operating temperature range. Each has a voltage of 12~V and capacity of 100~Ah. We connected eight batteries in series to achieve a total output of 96~V. This battery configuration provides a cumulative energy capacity of 9600~Wh. During polar day, the observation equipment can operate continuously, using solar energy, while during polar night, the batteries can sustain operation for approximately five days, based on the expected power consumption of 60~W.

Complete installation of the equipment required approximately eight hours of dedicated work on-site, with a few additional hours spent testing the installed instrument. The system was successfully powered up, and internal calibration procedures were completed before the expedition departed for Kunlun Station. Following 20 days of scientific work, the research team returned from Kunlun Station to Zhongshan Station, traversing over 100~km daily. On the second day of the return route, the team arrived at the designated experiment site for a scheduled operational window. Within scheduling constraints, field activities were limited to a critical 4-hour time frame for instrument status verification.

In the following two sections, we describe the design of the Antarctic global spectrum instrument, and the results of the experiment. 

\section{Instrument design} 
\label{sec:GSM}

The Antarctic global spectrum measurement instrument consists of an antenna, a wide-band receiver including the analog front end and the digital backend, and auxiliary systems which supply power. Aside from the usual amplifying channels, the receiver also needs a calibration subsystem, and a thermal control subsystem. The antenna uses an elliptical dipole configuration, with a smooth response needed for 21~cm global spectrum measurements, and also provides a natural enclosure for the receiver. The receiver is housed inside the elliptical cylinder dipole antenna. The received signals are amplified, digitized, and processed to produce the spectral data, which is stored on a hard drive for further analysis. To account for system drift over time, the instrument incorporates a self-calibrating mechanism that periodically alternates between calibrators and antenna according to a preset schedule. Unlike other global spectrum experiments, an instrument setup in Antarctica requires a special design for fully automatic operation, extreme cold environment, and the absence of maintenance for an entire year. 

\subsection{Antenna} 
\label{subsec: antenna}

The 21~cm global spectrum measurement generally requires a wide beam antenna with low chromaticity. According to the standard cosmological model, the 21~cm signal from the cosmic dawn era is probably redshifted to the frequency range of 50-100 MHz, which is the observing band. Correspondingly, the physical size is about half the wavelength, i.e., 2 meters.

The instrument is intended for deployment in the harsh environment of high plateaus in Antarctica, where the average altitude is 2350 meters above sea level, and up to 4000 meters near Dome A. The temperature during the polar day is around –30 \textdegree C, and during polar night can drop to –80 \textdegree C. The transportation and installation of a large, wide-band antenna in such conditions present significant logistical challenges. Therefore, the antenna design must incorporate a lightweight and modular structure for ease of transportation and installation.

Based on these considerations, the following design principles were established. \textbf{Low chromaticity} The antenna should maintain an almost invariant beam response across different frequencies within the observing band,  which is crucial for detecting the faint 21 cm signal in a wide-band system. \textbf{Lightweight and integrated design} Given the harsh environment and difficulty of working condition, the antenna should be lightweight and easy to transport and assemble. With this in mind, we designed the antenna to also house the receiver electronics, so that it is unnecessary to have another, separate, electronics box; this also avoids a complicated balun structure. \textbf{Electromagnetic self-shielding} As the antenna also serves as the enclosure of the electronics, it must have robust electromagnetic self-shielding to prevent interference from the receiver and power supply, where numerical digital circuits are necessitated.

An elliptical dipole antenna satisfies these requirements. A planar dipole antenna was positioned horizontally above ground at a height of approximately one-quarter the central wavelength, with its maximum gain directed toward the zenith. Instead of a planar blade dipole, we adopted a pair of elliptical cylinders with finite thickness, the inside of which can also house the receiver. 

\begin{figure}[htbp]
\centering
\includegraphics[width=0.5\columnwidth]{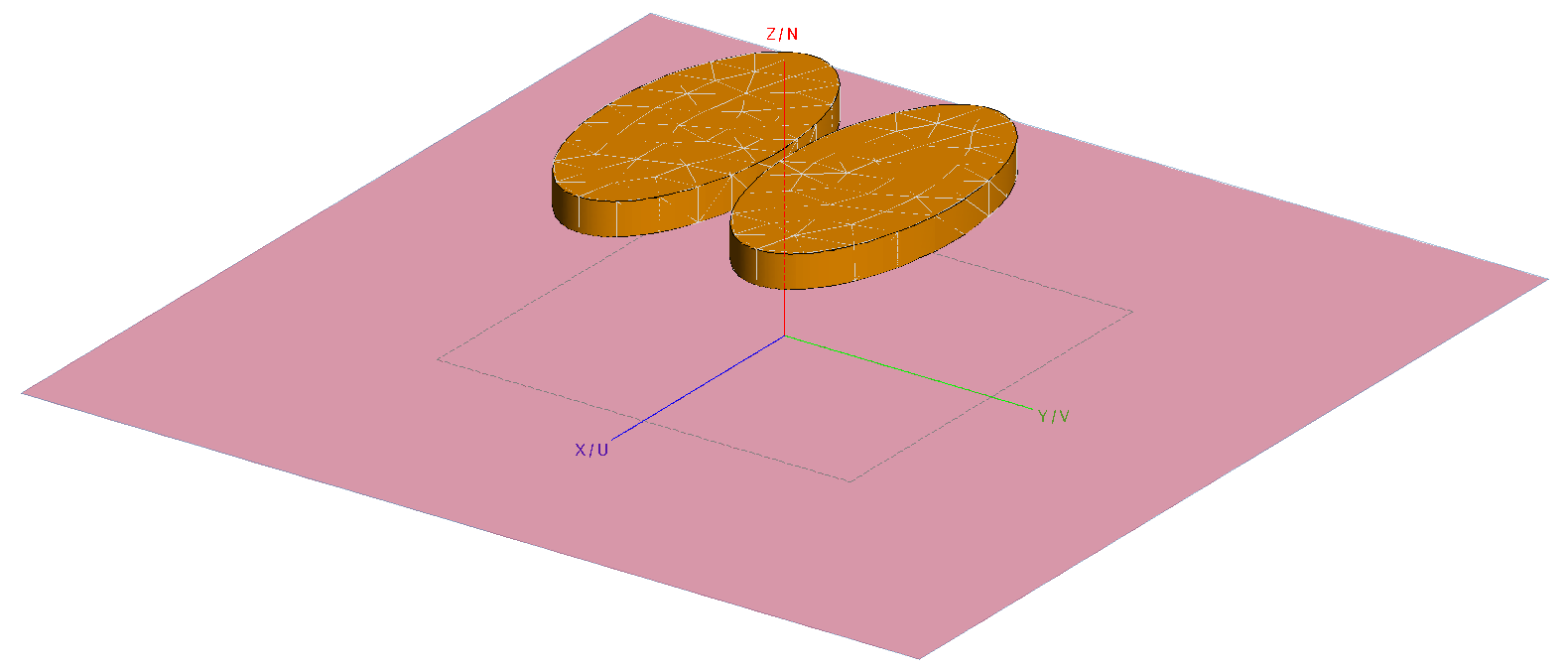}
\includegraphics[width=0.5\columnwidth]{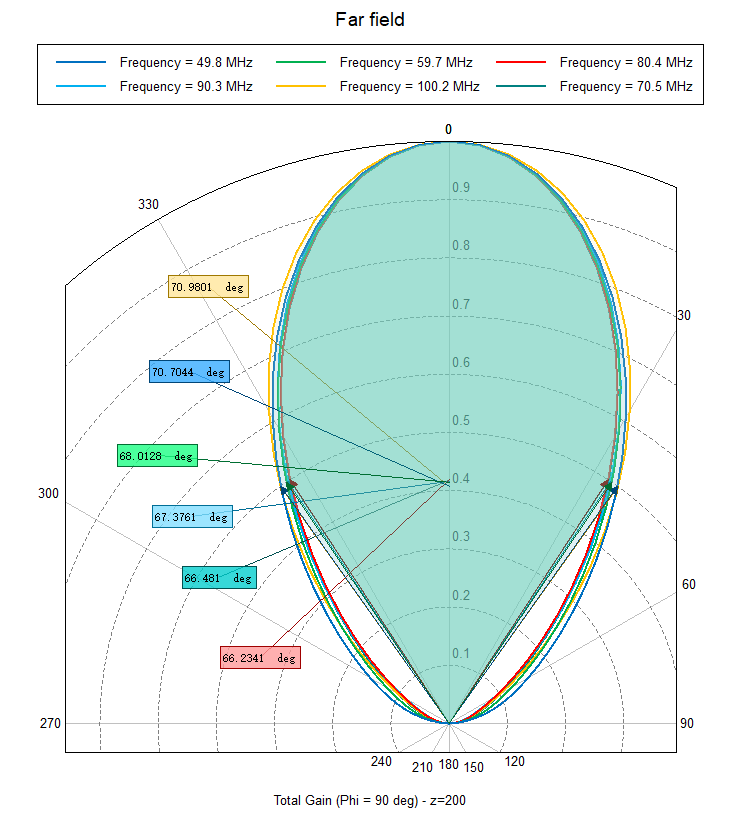}
\caption{Top: 3D-rendered model of the elliptical antenna. Bottom: The simulated far-field beam of the antenna for 6 frequency points in the range of 50~--~100~MHz, and their corresponding 3~dB beamwidths.}
\label{fig:antenna}
\end{figure}

The top panel of Fig.~\ref{fig:antenna} shows the elliptical dipole antenna model. The elliptical shape was chosen due to its smooth edge transitions, which allows for the uniform flow of current along the sides of the antenna. This results in a beam that points toward the zenith, and varies very little across the observing frequency band, as well as ensuring good impedance-matching (shown in the bottom panel of Fig.~\ref{fig:antenna}). 

For optimal performance, this antenna is supported by a non-conducting frame above the ground. Given that wind speeds during Antarctic blizzards can reach up to 70~m/s, a supporting frame with exceptional strength and rigidity is required to maintain stability without affecting the electrical properties of the antenna. To meet these requirements, the support frame is constructed primarily from glass fiber reinforced plastic (GFRP), which offers a tensile strength of approximately 3500~MPa, while also exhibiting lower electrical conductivity, thereby minimizing any interference with electrical performance of the antenna. 

\begin{figure}[htbp]
\centering
\includegraphics[width=0.5\columnwidth]{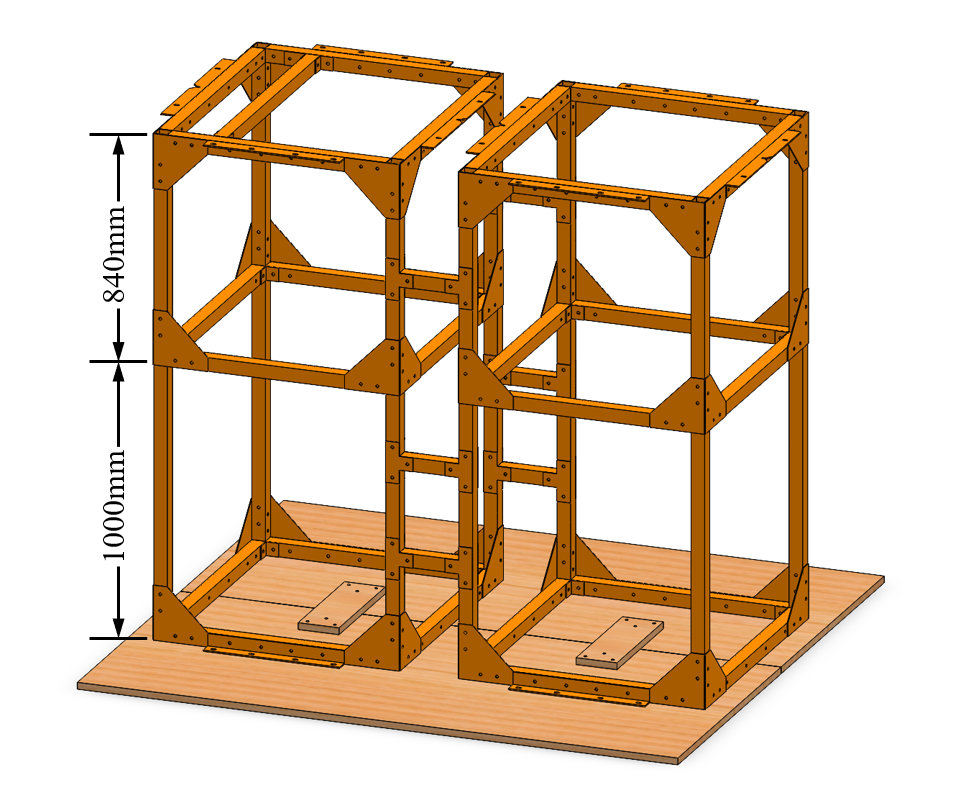}
\caption{3D rendering of the antenna support frame.} 
\label{fig:antenna bracket}
\end{figure}

The overall structure of the support frame (shown in Fig.~\ref{fig:antenna bracket}) is divided into two symmetrical sections, left and right, connected at the base by large wooden planks. The top surface of the frame serves to support the antenna. Since the left section of the frame houses the receiver system chassis, additional square GFRP tubes are added to prevent deformation.
The total height of the support frame is 1840~mm, with 840~mm of the upper section exposed above the ice surface to ensure optimal antenna performance. The lower portion, 1000~mm in height, is buried in the ice to provide stability against wind. This design prevents the antenna, which has a large surface area, from tipping over in extreme weather conditions. Additionally, the wooden planks at the base of the support frame prevent the entire antenna structure from sinking into the snow or ice over time. This structural design ensures both mechanical stability and long-term functionality of the antenna system in the harsh Antarctic environment.

\subsection{Receiver} 
\label{subsec: receiver}
\begin{figure*}[htbp]
\centering
\includegraphics[width=\columnwidth]{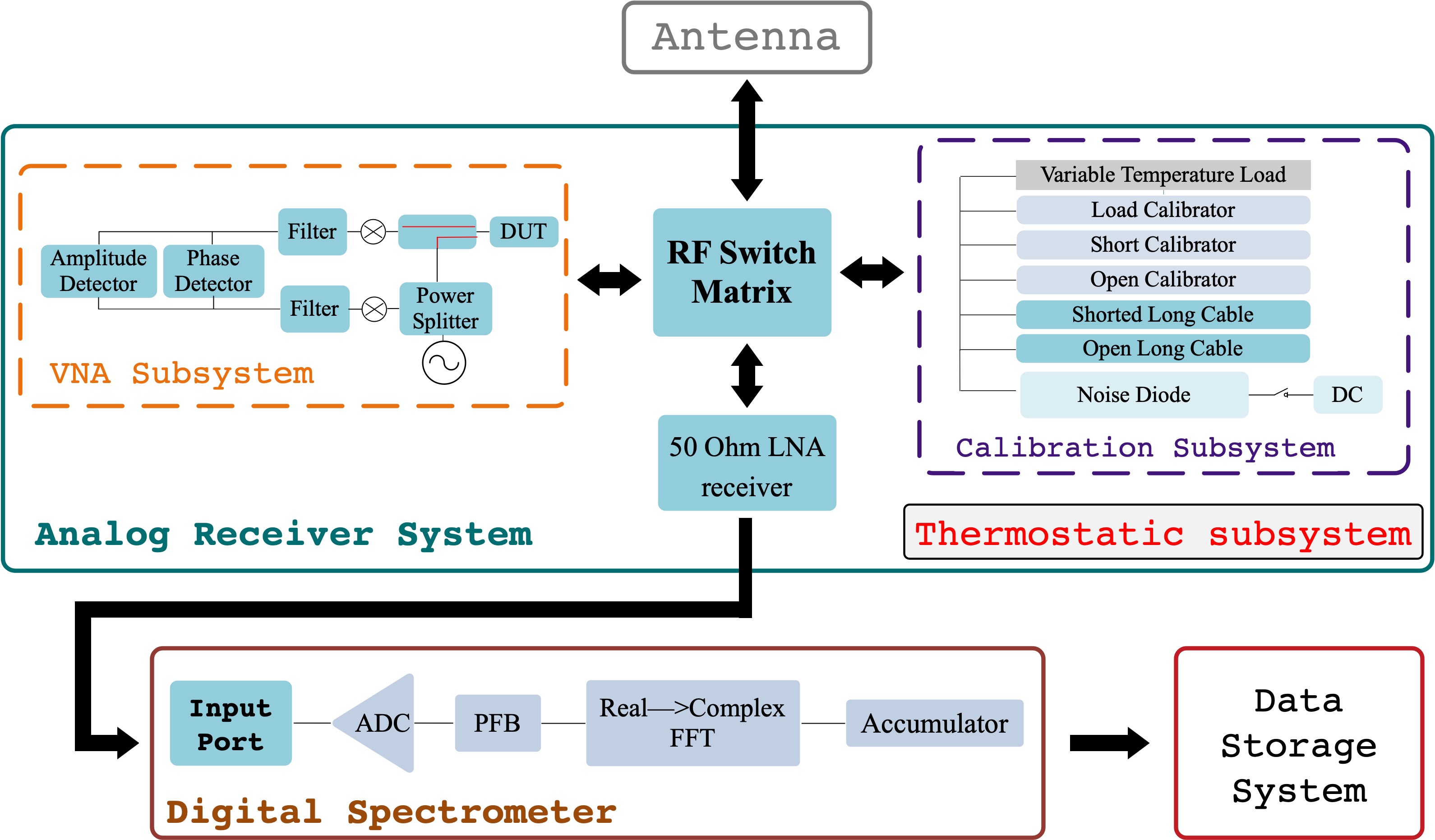}
\caption{Receiving system schematic, showing the analog receiving system, thermostatic subsystem for the analog receiver, digital spectrometer, and data storage system.}
\label{fig:schematic}
\end{figure*}

The overall structure of the receiving system is shown in Fig.~\ref{fig:schematic}. The signal from the receiving antenna is sent to the analog receiver system. The analog or calibration signal is sent to the digital spectrometer, where it is digitized and processed, then saved in the data storage system. A more detailed description of our global spectrum receiving system can be found in ~\citet{wu2024}. 

\subsubsection{Analog Receiver}
The analog receiver comprises several subsystems and components. The low-noise amplifier (LNA), and associated filters, to amplify the signal; the calibration subsystem and vector network analyzer (VNA) subsystem, which are used for calibration; the radio frequency (RF) switch matrix, which switches the input between antenna, calibrator, and VNA, and; the thermostatic subsystem, which maintains the temperature of the receiver.

\begin{figure}[htbp]
\centering
\includegraphics[width=0.5\columnwidth]{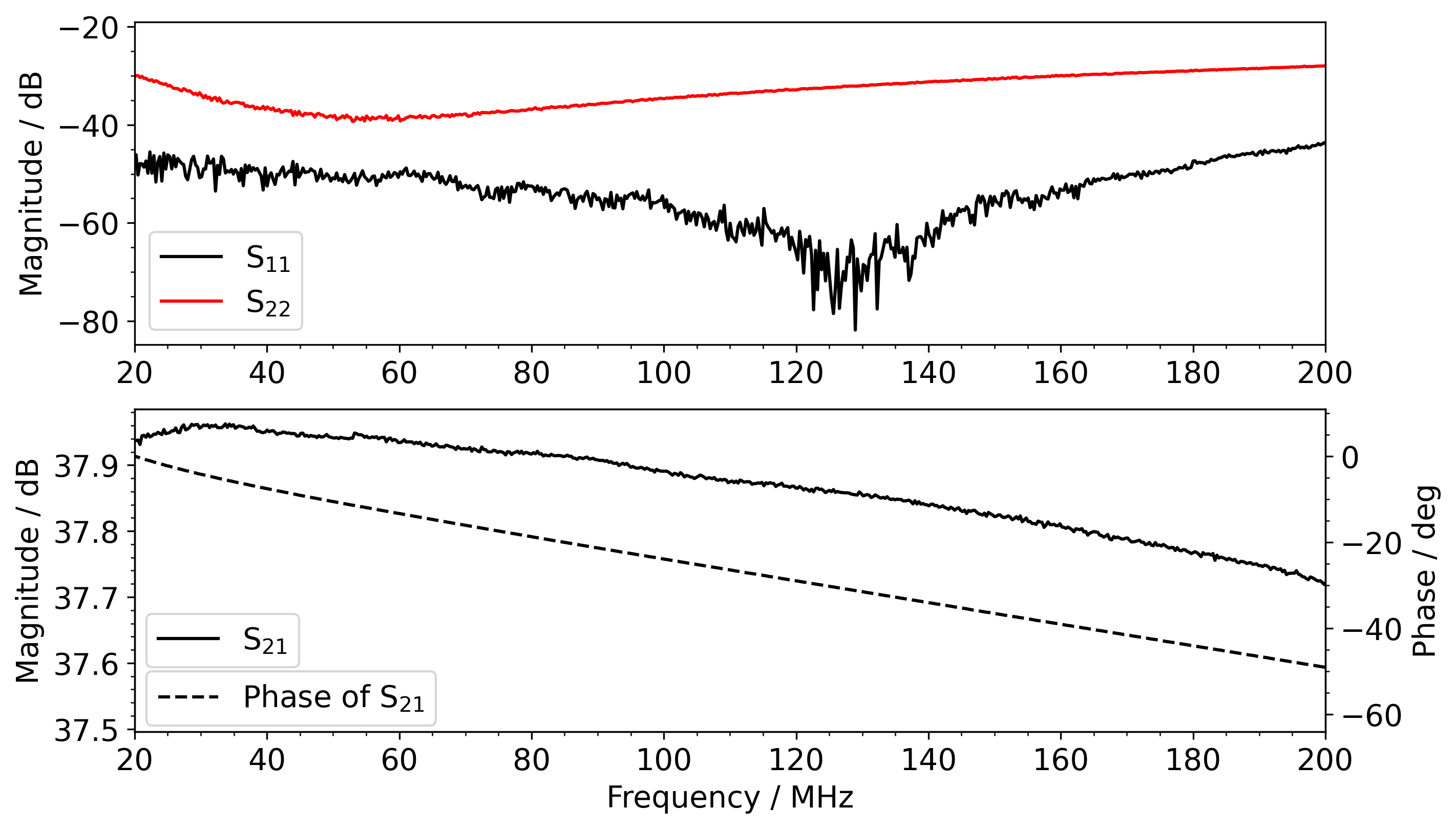}
\caption{Receiver measurement results. The top panel shows the receiver reflection coefficients, $S_{11}$ and $S_{22}$, and the bottom panel shows the transmission coefficient, $S_{21}$.}
\label{fig:receiver measurement}
\end{figure}

{\bf LNA}
The custom-made LNA is optimized for low noise, and low reflection coefficients for both input and output, $S_{11}$ and $S_{22}$, which effectively reduce the reflected noise between antenna and receiver. The $S_{11}$, $S_{22}$, and gain of the LNA are plotted in Fig.~\ref{fig:receiver measurement} as a function of frequency. Measurements of the LNA show an average $S_{11}$ of --45~dB, and the average $S_{22}$ is --30~dB. The LNA offers a typical  gain of 37~dB with a flatness of 0.2~dB.

\textbf{Calibration subsystem}
The calibration subsystem uses a variable temperature load to calibrate the temperature drift, a noise diode, an open long cable, a shorted long cable and the VNA calibrators. The RF switch matrix is employed to control signal paths, ensuring efficient transitions between calibration and measurement modes.

This receiver channel can be switched to one of six inputs: the antenna for observation of the sky signal; a noise diode to calibrate the relative change of the system gain; the variable temperature load, which can be heated to 400~K, for absolute calibration, with an accuracy exceeding 0.1~K; the calibrators for calibration of VNA characteristics; an open long cable; and a shorted long cable used to solve the noise wave parameters, $T_u$, $T_c$, $T_s$, and $T_0$ of the system.

\textbf{VNA subsystem}
A VNA is used for precise impedance calibration of both the antenna and receiver. Impedance mismatch between the antenna and receiver can lead to signal reflections and standing waves, which introduce additional noise. During self-calibration, the VNA subsystem facilitates real-time measurement of the antenna and the receiver impedance. This real-time measurement enables a more precise determination of the system response.

\textbf{Thermostatic subsystem}
A 21~cm global spectrum measurement requires high stability. To achieve this, the analog front end is integrated with the thermostatic subsystem. Active components, including amplifiers, are housed within small, precision-controlled enclosures that use a 4~$\times$~4~cm thermoelectric cooler to maintain the temperature to within $\pm$0.1\textdegree C. The thermostatic subsystem uses a purely analog circuit to avoid possible electromagnetic interference (EMI). It is controlled by an analog Proportional-Integral-Differential circuit. With components exhibiting minimal temperature drift, the front-end gain variation is maintained within 0.03~dB. We set the equilibrium temperature at --30\textdegree C, since the daytime temperature in polar day is below --20\textdegree C, and the “nighttime” temperature is about --30\textdegree C. When the thermostatic subsystem is cooling, the peak power consumption is less than 30~W, and in a stable state, the power consumption is 3~W.

\subsubsection{Digital spectrometer} 
The digital spectrometer is designed for high-precision, wide-bandwidth sampling, characterized by low power consumption (approximately 20~W) and robust thermal stability. 
The digital board employs a Xilinx Kintex7 high-performance Field-Programmable Gate Array and a 12-bit, 500~MSPS TI~ADS5407 chip to provide high-dynamic range sampling. The system achieves a frequency resolution of 8192 channels, thereby ensuring adequate frequency resolution and dynamic range. It has been tested to function effectively at temperatures as low as --40\textdegree C. 

\begin{figure}[htbp]
\centering
\includegraphics[width=0.5\columnwidth]{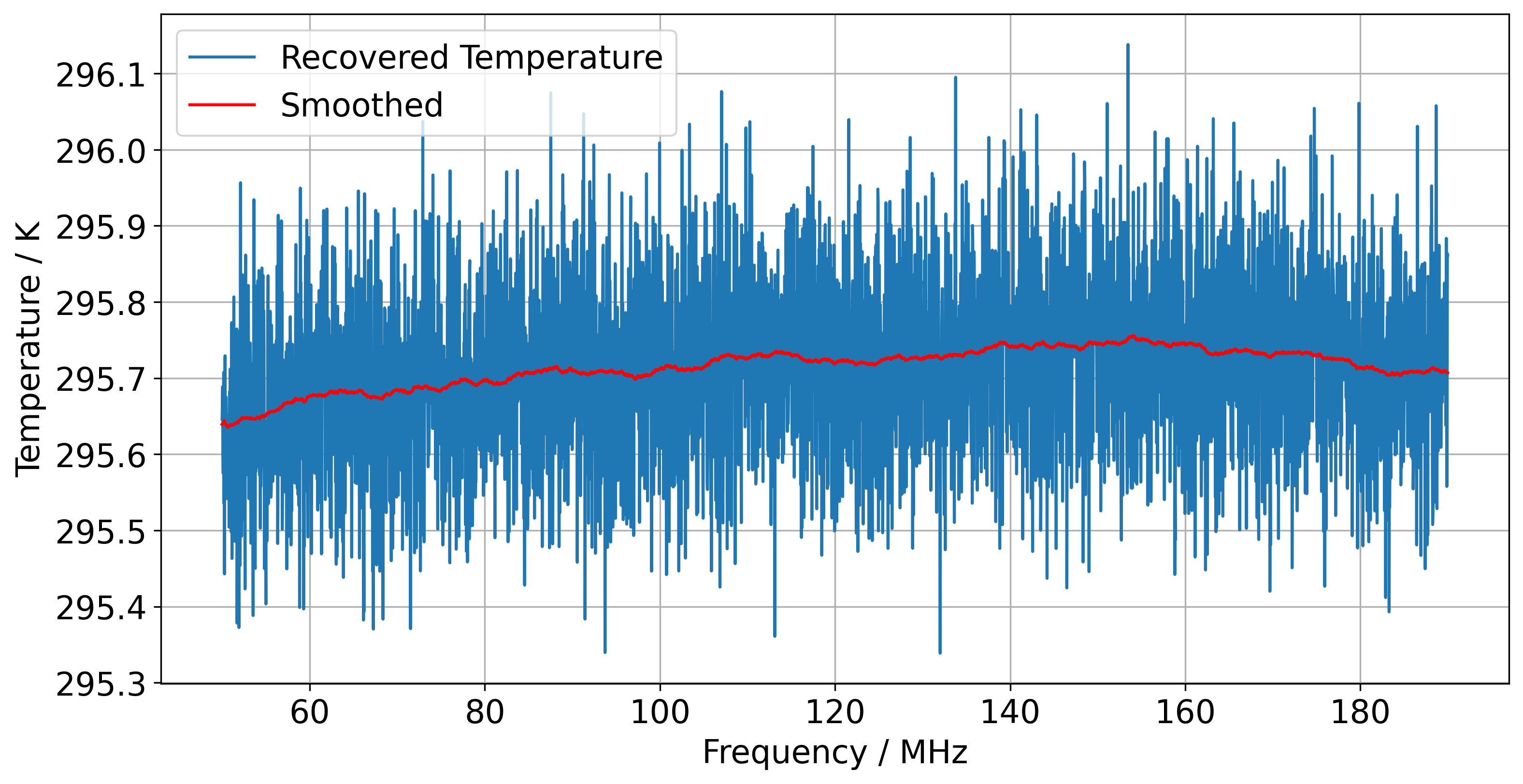}
\caption{Calibration results. The blue curve shows the recovered temperature of an ambient antenna simulator. The red curve shows the smoothed value.}
\label{fig:receiver calibration in lab}
\end{figure}

The receiver system integrates self-calibration mechanisms to ensure high sensitivity and accuracy.  Fig.~\ref{fig:receiver calibration in lab} shows the calibration result when an antenna simulator (load) is connected. After an integration time of two hours, it can measure the load physical temperature (in this case equal to ambient temperature) to within 0.1~K. We estimate that our instrument has a sensitivity of 0.1~K across the entire observation band. The instrument specification is shown in Table ~\ref{tab:specification}.

\begin{table}[!ht]
    \renewcommand{\arraystretch}{1.3}
    \centering
    \caption{Instrument technical specifications.}
    \label{tab:specification}
    \begin{tabular}{c | c}
        Specifications & Value \\ \hline
        Frequency Range & 30-200~MHz \\ 
        Gain & 70~dB \\ 
        Gain Stability & 0.01~dB \\ 
        Out-of-band Rejection & 60~dB \\ 
        Sensitivity@30-200~MHz & 0.1~K \\ 
        Sensitivity@50-100~MHz & 0.04~K \\ 
        Sampling Rate & 500~MSPS \\ 
        Sampling Precision & 12~bit \\ 
        Frequency Resolution & 61~kHz \\ 
    \end{tabular}
\end{table}

\subsection{Receiver chassis} 
\label{subsec: receiver chassis}
\begin{figure}[htbp]
\centering
\subfigure[Top view]{
\includegraphics[width=0.5\columnwidth]{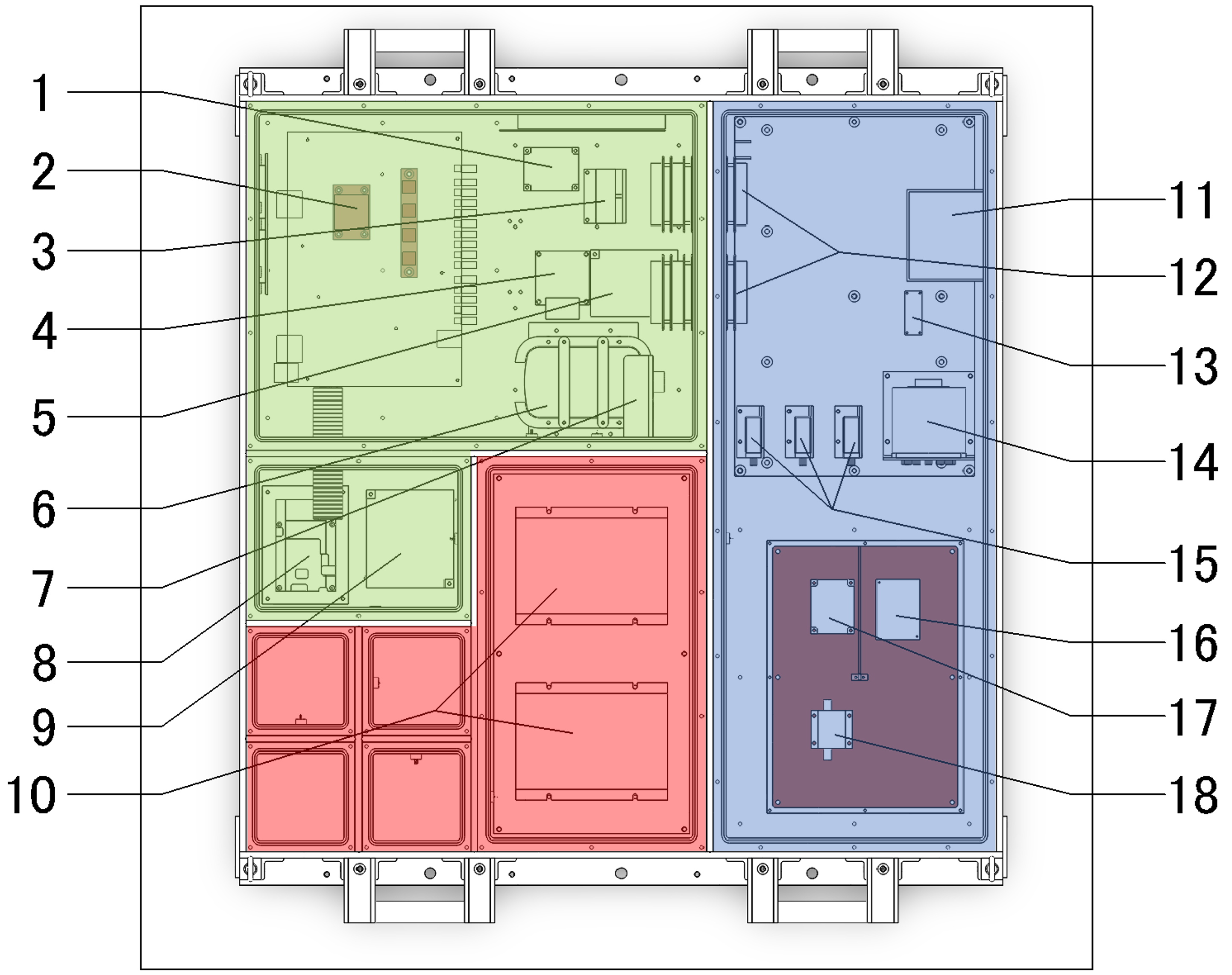}
}
\quad
\subfigure[Side view]{
\includegraphics[width=0.5\columnwidth]{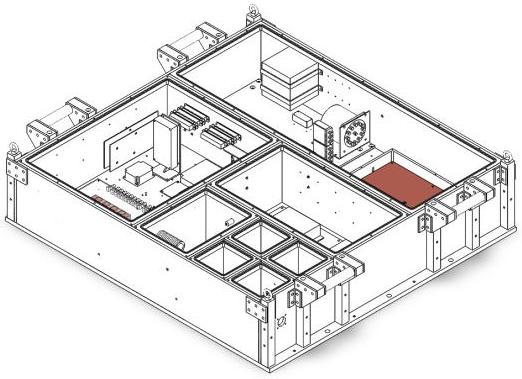}
}
\caption{Three-dimensional diagram of the receiver chassis. The digital section is shown in the green shaded area, comprising: 1. Oscillator. 2. Digitizer. 3. USB hub. 4. Serial port hub. 5. Temperature controller. 6. VNA. 7. Fan. 8. Microcomputer. 9. Temperature controller. The power supply section is denoted in the red shaded area: 10. DC-DC converter. The analog section is denoted in the blue shaded area: 11. Variable temperature load. 12. 40-pin connectors. 13. Solid state relay. 14. 1P12T RF switch. 15. 1P2T RF switch. 16. Thermostatic chamber. 16-1. Noise source. 16-2. Low-noise amplifier. 16-3. Second-stage amplifier.}
\label{fig:receiver chassis}
\end{figure}

Both the analog front end and digital backend of the receiver, along with the power supply unit, are housed within a chassis integrated into the body of the elliptic dipole antenna.
 The components of the analog section include variable temperature loads, RF switch, RF switch logic control board, solid state relays, and a compact thermostat enclosure for the LNA, noise source, and the second-stage amplifier. The digital section comprises digital acquisition boards, frequency oscillator source, temperature measurement unit, USB hub, serial hub, VNA, temperature controllers, and a microcomputer. The overall layout for the receiver system is shown in Fig.~\ref{fig:receiver chassis}(a). The digital section is denoted in green, the analog section in blue, and the power supply section in red.

The receiver chassis requires careful electromagnetic compatibility design to ensure effective shielding for each section. Furthermore, the RF devices are sensitive to temperature variations, necessitating a dedicated thermostatic design. The chassis contain three separate compartments, housing the digital, analog, and power supply sections, as shown in Fig.~\ref{fig:receiver chassis}. This compartmentalized design mitigates EMI by the digital and power supply component on the sensitive analog component. The arrangement of each component within the chassis is optimized based on size and functionality. The physical dimensions of the receiver case are 700 mm~$\times$~700 mm~$\times$~180 mm. There are four cavities between the switching power supply and the electronic equipment, and multistage low-pass EMI filters are inserted into these cavities. The external connections of the whole chassis are limited to three cables. One coaxial line is for the antenna feed, which transmits the signals received by the antenna to the analog section. A second coaxial line is for external data transmission, allowing for external communication with the microcomputer of the system. Lastly, there is a cable to the power supply section from the battery.  

\begin{figure}[htbp]
\centering
\includegraphics[width=0.5\columnwidth]{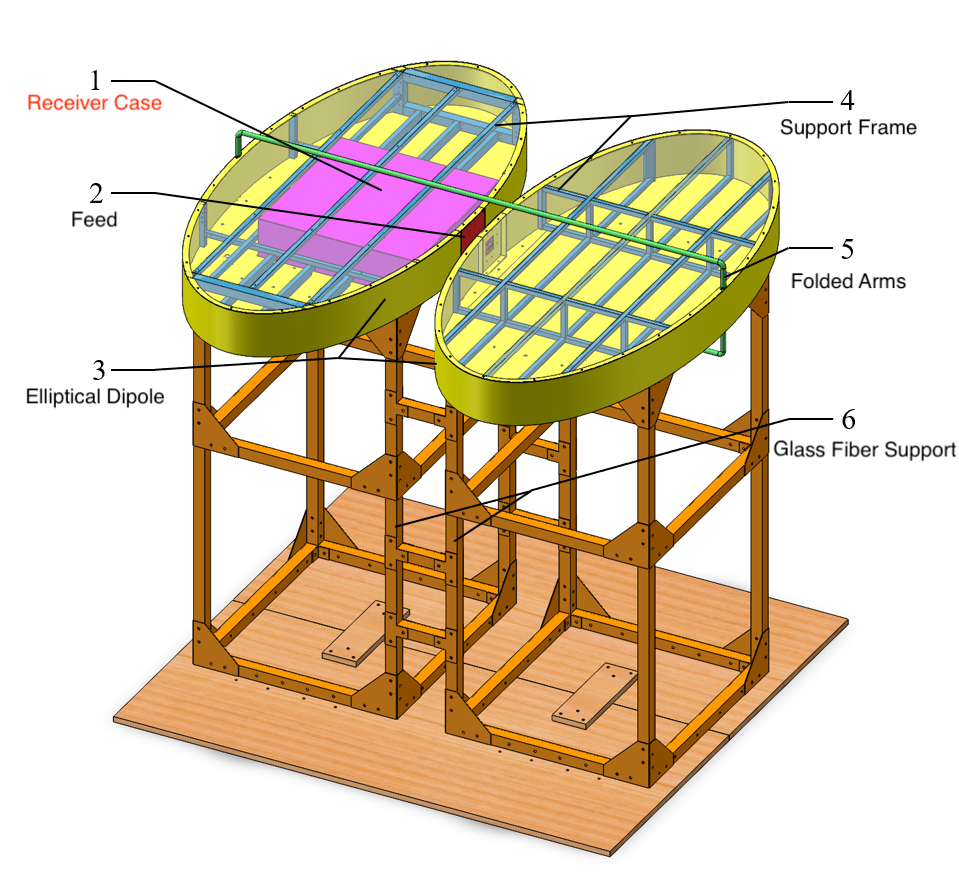}
\caption{3D rendering of the Antarctic global spectrum measurement instrument}
\label{fig:three-dimensional diagram}
\end{figure}

The antenna body must be rigid, to maintain its shape and avoid deformation, while also needing to be lightweight to facilitate transportation and installation in the harsh Antarctic environment. Our design is based on a thin shell with internal reinforcement. The overall structure of the antenna is shown in Fig.~\ref{fig:three-dimensional diagram}. The receiver case is placed in one of the dipoles, further reinforced with crossbars.

\subsection{Power supply} 
\label{subsec: power}

\begin{figure}[htbp]
\centering
\subfigure[Power supply schematic.]{
\includegraphics[width=0.5\columnwidth]{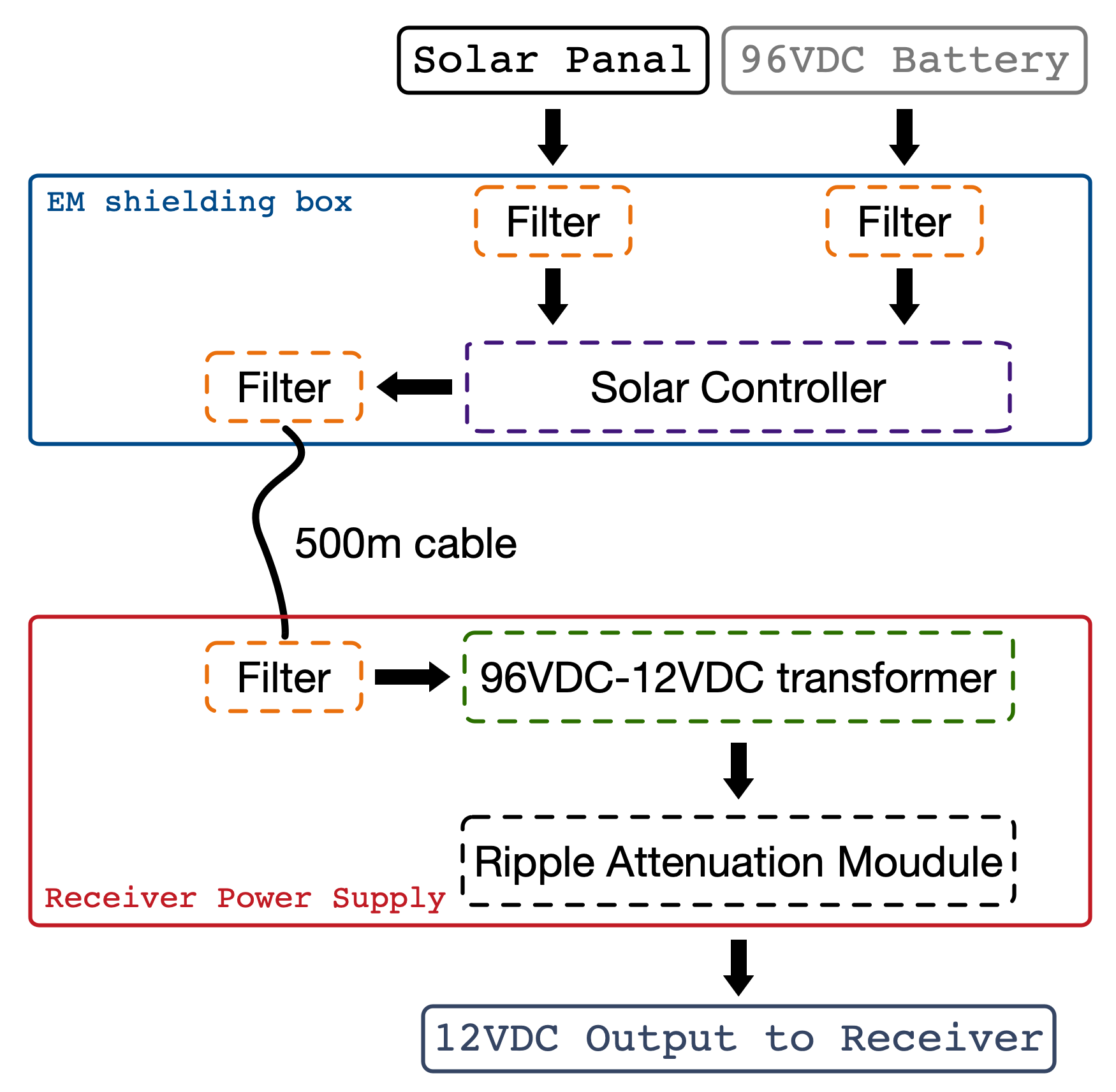}
}
\quad
\subfigure[Solar controller enclosure.]{
\includegraphics[width=0.5\columnwidth]{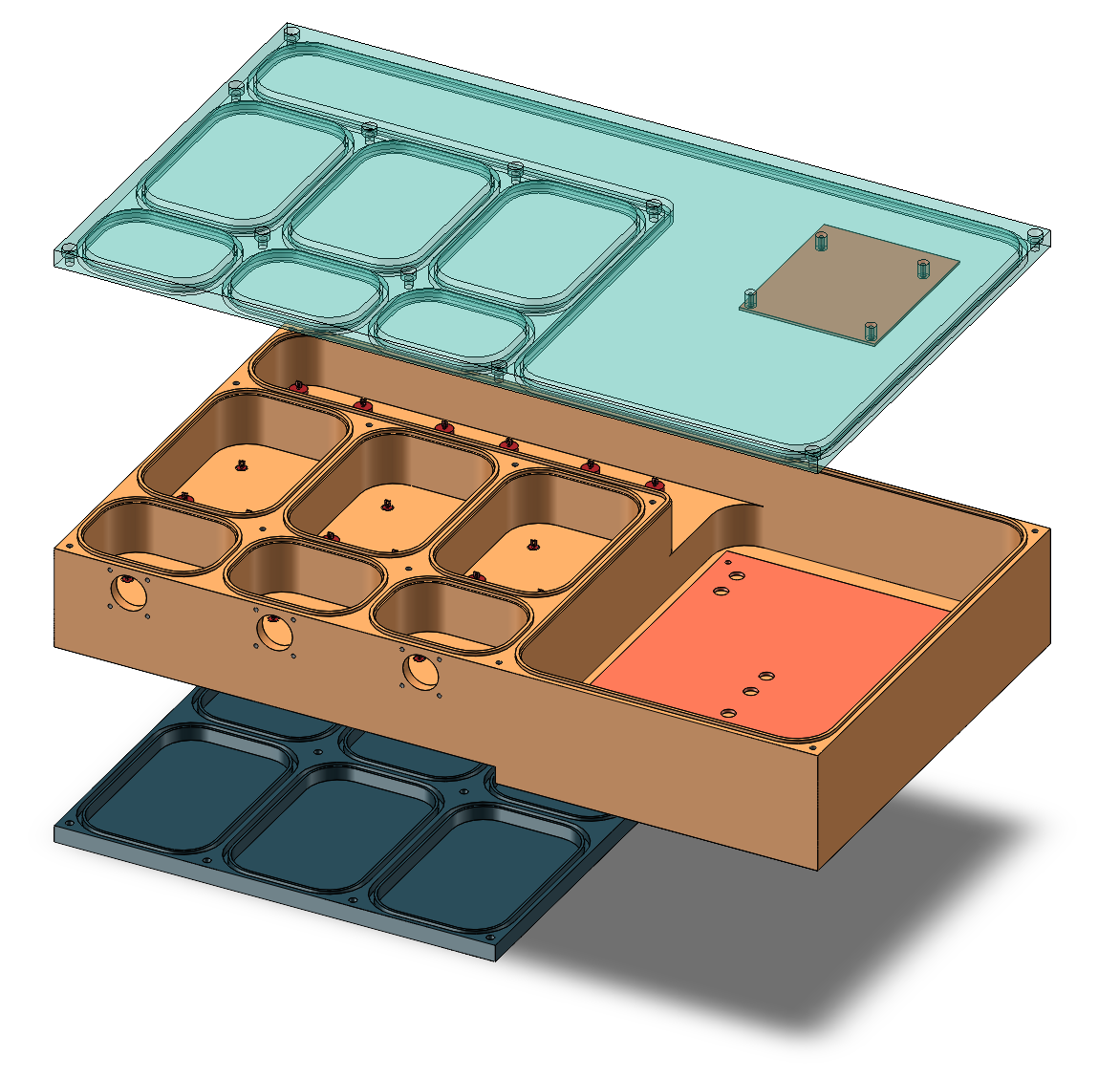}
}
\caption{Power supply design.}
\label{fig:power supply}
\end{figure}

Laboratory tests show that the power consumption of the Antarctic global spectrum measurement instrument is approximately 60~W in common observation mode. We also deployed an instrument of similar design in Xinjiang, China, with operating results indicating that the power consumption is nearly 60~W in either winter  or summer (when the average temperatures are --20\textdegree C and 30\textdegree C, respectively). To provide energy, the solar panels charge the batteries during polar day, via a solar controller, which subsequently provides power to the device. During polar night, the device enters a dormant state upon battery depletion and is programmed for automatic activation at the onset of polar day to resume observations. The proposed power supply configuration for the instrument is illustrated in Fig.~\ref{fig:power supply}(a). The solar controller features three inputs/outputs, corresponding to the solar panel, battery, and load. 

At the site, which has a latitude of $\sim$80\textdegree S, the Sun does not rise more than 35\textdegree\ above the horizon, so we installed the solar panels in a vertical orientation. The standard output voltage for each panel is 50~V with the panels connected in series, in groups of three, yielding a combined output of 150~V. The selected solar controller operates at a mean charge voltage of 113.6~V (maximum 230~V) to charge the batteries.

A significant challenge associated with this energy supply scheme is that both the solar controller and the DC-DC unit for voltage conversion produce low-frequency EMI, which could severely affect our experiment. To mitigate this issue, electromagnetic shielding enclosures have been designed to house any components prone to generating electromagnetic radiation, as shown in Fig.~\ref{fig:power supply}(b). Additionally, the energy supply unit is positioned 500~m away from the antenna, to minimize the potential impact of EMI. 

Within the power supply section of the system chassis, the current first passes through a filter, after which a DC-DC module converts the high voltage of 96~V down to the 12~V required by the equipment housed within the chassis. A ripple suppressor is applied to further refine the output current to pure DC. For active components within the analog front end, such as the LNA and noise calibration sources, additional linear voltage regulators are employed for voltage conversion at each individual component.

\subsection{Design and implementation for extreme environment}
\label{subsec: low-temperature design}

The environmental temperature of the Antarctic inland is approximately --40 to --30\textdegree C during polar day, and the lowest temperature can reach --70\textdegree C during polar night. Providing energy during the polar night is difficult, because no solar power is available during the long polar winter when the Sun is below the horizon for several months. We therefore design our instrument to be able to survive unpowered during polar night, but still operate when power is restored. To ensure our instrument functions normally in the extreme Antarctic environment, considerable design, improvement, testing, and experimental work have been done. 

\textbf{Material selection} Many of the modules in the instrument are custom-designed, including the LNA, frequency source, and RF filters. These are constructed using military-grade chips, with operating temperature ranges of --55 to +150\textdegree C, suitable for the extreme Antarctic environment. For the cables, we used ultra-flexible silicone rubber cables, with a low-temperature tolerance of --60\textdegree C. For the batteries, we selected a variety with the highest available charge and discharge efficiency at low temperatures (see Section~\ref{subsec: power}).
\textbf{Reinforcement} Various commercial products used in our experiment, such as the microcomputer, solar controller, and serial port controller, were reinforced by replacing some of their components. For example, we replaced electrolytic capacitors, which are prone to damage at low temperatures, with solid state capacitors. 
\textbf{Thermal insulation} Some modules only have good performance at standard temperatures, such as microwave switches, mechanical calibrators, hard disks, and batteries. We have taken thermal insulation measures for these modules, wrapping the cavity where these modules are placed with special thermal-insulating cotton. 
\textbf{Laboratory test} We conducted low-temperature tests on most of the modules in a high-and-low-temperature chamber. These modules were tested under operating conditions, with the chamber set at --50\textdegree C. To verify that these modules will not be damaged during polar night, specifically at a temperature of --70\textdegree C, we powered off the modules, then set the chamber to --70\textdegree C. When the temperature rose to --50\textdegree C, we powered them on again to check whether they could be successfully rebooted. 
\textbf{Field experiment.} We have also conducted system-level tests by deploying an identical set of equipment in the electromagnetic quiet zone at Hongliuxia Station in Xinjiang. The temperature there during winter can drop below --20\textdegree C. This is still much higher than Antarctic temperatures, but is suitable for some preliminary testing.

\section{Results}
\label{sec:results}

\subsection{RFI measurement}
We measured the RFI along the route from Zhongshan Station ($69^\circ22^\prime24.76^{\prime\prime}$S, $76^\circ22^\prime14.28^{\prime\prime}$E), to Kunlun Station ($80^\circ25^\prime 01^{\prime\prime}$S, $77^\circ06^\prime 58^{\prime\prime}$E). At each stop of the day, RFI measurements were taken at a distance of approximately 2~km from the camp of the convoy.

\begin{figure}[htbp]
\centering
\includegraphics[width=0.5\columnwidth]{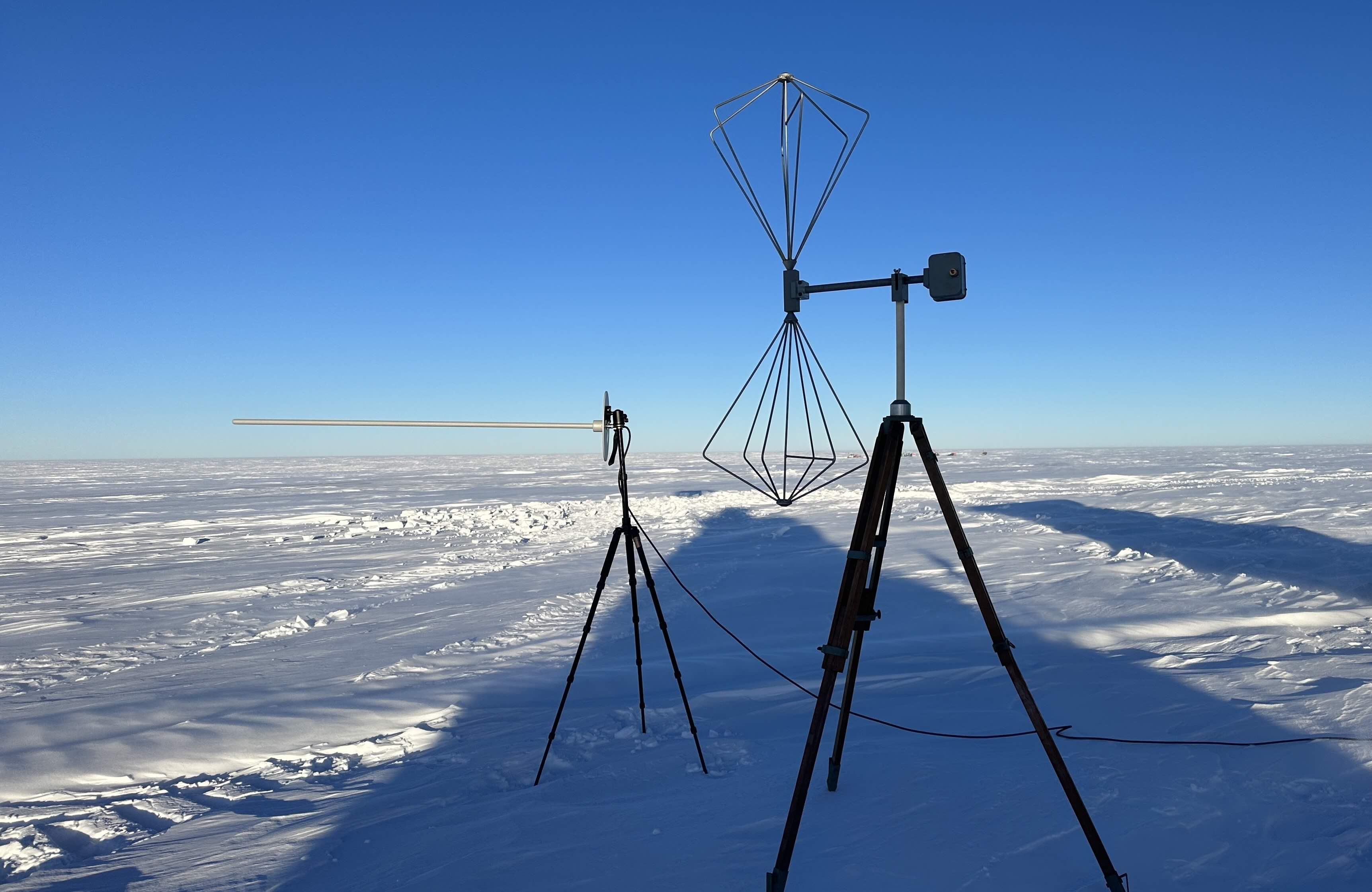}
\caption{RFI measurement antennas.} 
\label{fig:RFI antenna}
\end{figure}

The RFI is measured with a spectrum analyzer, using a rod antenna for lower frequencies (10~kHz~--~30~MHz), and a biconical antenna for higher frequencies (30~MHz~--~300~MHz), as shown in Fig.~\ref{fig:RFI antenna}. The RFI is measured to a spectral resolution of 1~kHz, in three orthogonal directions by each antenna: vertical, north-south horizontal, and east-west horizontal.

\begin{figure}[htbp]
\centering
\includegraphics[width=\columnwidth]{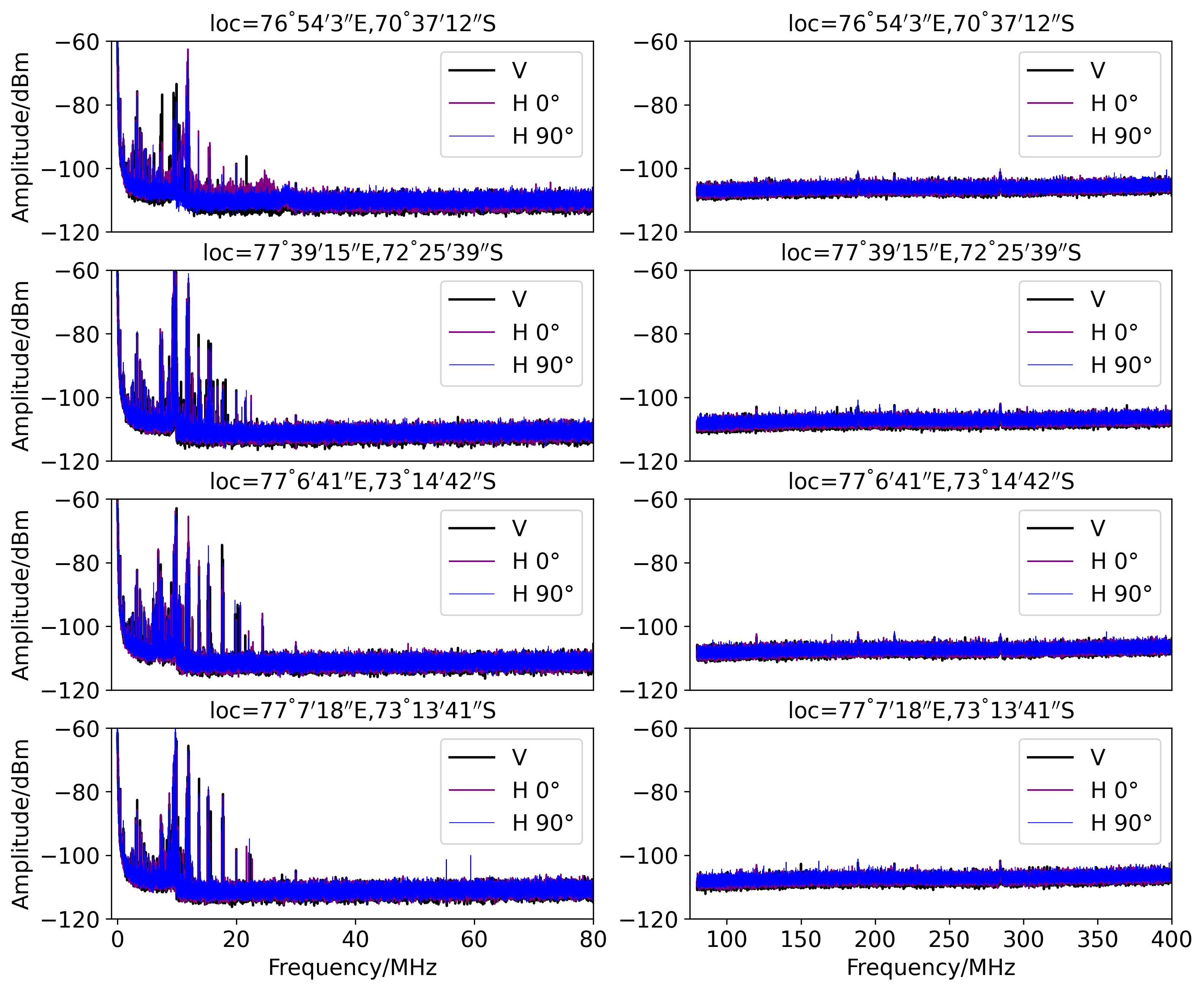}
\caption{RFI measurement results. Left: 0.1~--~80~MHz. Right: 80~--~400~MHz. All traces are in the ``Max Hold' mode.} 
\label{fig:RFI measurement}
\end{figure}

Unfortunately, measurements can only be taken near the convoy camp, and for safety reasons, convoy activities inevitably generate significant RFI, from sources such as the tractor engine, computer and communication equipment, and other instruments. Consequently, measured results may overestimate the true RFI values. 
Major interference, measured in four typical locations, is distributed at frequencies below 30~MHz, while there is little RFI in the 30~--~400~MHz region (see Fig.~\ref{fig:RFI measurement}). Even with the convoy present, the RF environment is excellent in the 30~--~400~MHz band. 

\subsection{Ground effect}
As noted in Section~\ref{sec:intro}, ground effects have significant impacts on high-precision global spectrum measurements. In most locations, soil is a poor conductor, and its conductivity and dielectric constant change with temperature and humidity. The conductivity of soil is also highly dependent on its water content, which increases substantially below the water table line and in areas with conductive minerals, so that strong reflected waves can be generated in underground layers, forming standing waves with the antennas located at the surface. These difficult-to-measure changes ultimately cause systematic errors, producing complex and difficult-to-identify effects.
To address this challenge, most experimental setups employ a metal ground plane—typically a high-conductivity metallic mesh installed around the detection antenna—to mimic an ideal conductive plane for simplified analysis. However, the physical size of the metallic ground plane and unavoidable environmental couplings cause constraints, so complete elimination of such interference remains unachievable.

Antarctica is covered with an ice cap with an average thickness of 2~km, reaching 3~km in some places. The surface of the ice cap is flat, so that no strong reflection or scattering occurs, except for the interface between the ice surface and the air. In our frequency band of interest, the attenuation of electromagnetic waves within the Antarctic ice sheet is rapid. Simulations show that transmitted waves in the ice sheet are almost completely absorbed by a 100-meter-thick ice layer, so reflections from lower layers do not have a significant impact. Reflection occurring at the surface can be precisely simulated using electromagnetic simulation software, and it serves as part of the antenna response. In addition, Antarctica temperatures are extremely low, and the ice and snow surface are stable, which can alleviate or even avoid problems caused by changes in dielectric constant and conductivity on the ground.

\begin{figure*}[htbp]
\centering
\includegraphics[width=\columnwidth]{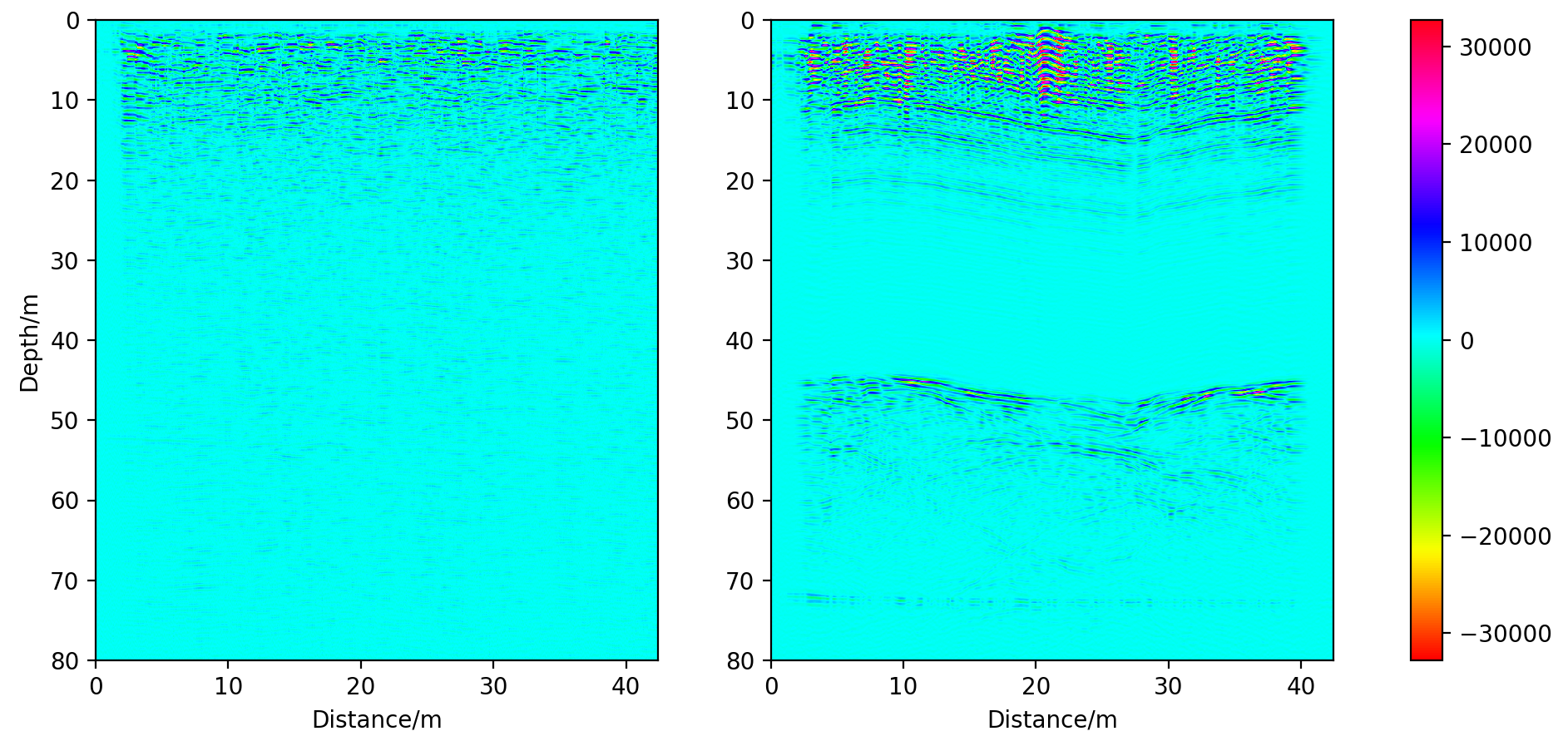}
\caption{GPR measurement results. Left: The Antarctic inland region, where the thickness of the ice cap is over 2~km. Right: a site near Zhongshan Station, where the thickness of the ice cap is approximately 40~m. The color bar shows the strength of the reflected wave (ADC units).
}
\label{fig:GPRresult}
\end{figure*}

To further investigate the effect of the ice sheet, we used GPR to probe the ice layers during the expedition. The GPR unit was moved on the ice surface along a straight line while beaming radar waves downward. Fig.~\ref{fig:GPRresult} shows the ice sheet reflections. 

Most reflections occur near the ground surface. In the inland ice cap, we do not observe significant structure. The reflected waves are stronger near the surface, but that is mostly due to the fact that the incident wave is stronger there. These randomly reflected waves will not generate spectral features that could be confused with the 21~cm signal. By contrast, the seashore figure shows clear layers, which could generate confusing spectral features.  

\subsection{In situ observation}
\label{subsubsec:observation}

\begin{figure}[htbp]
\centering
\subfigure[Antenna]{
\includegraphics[width=0.5\columnwidth]{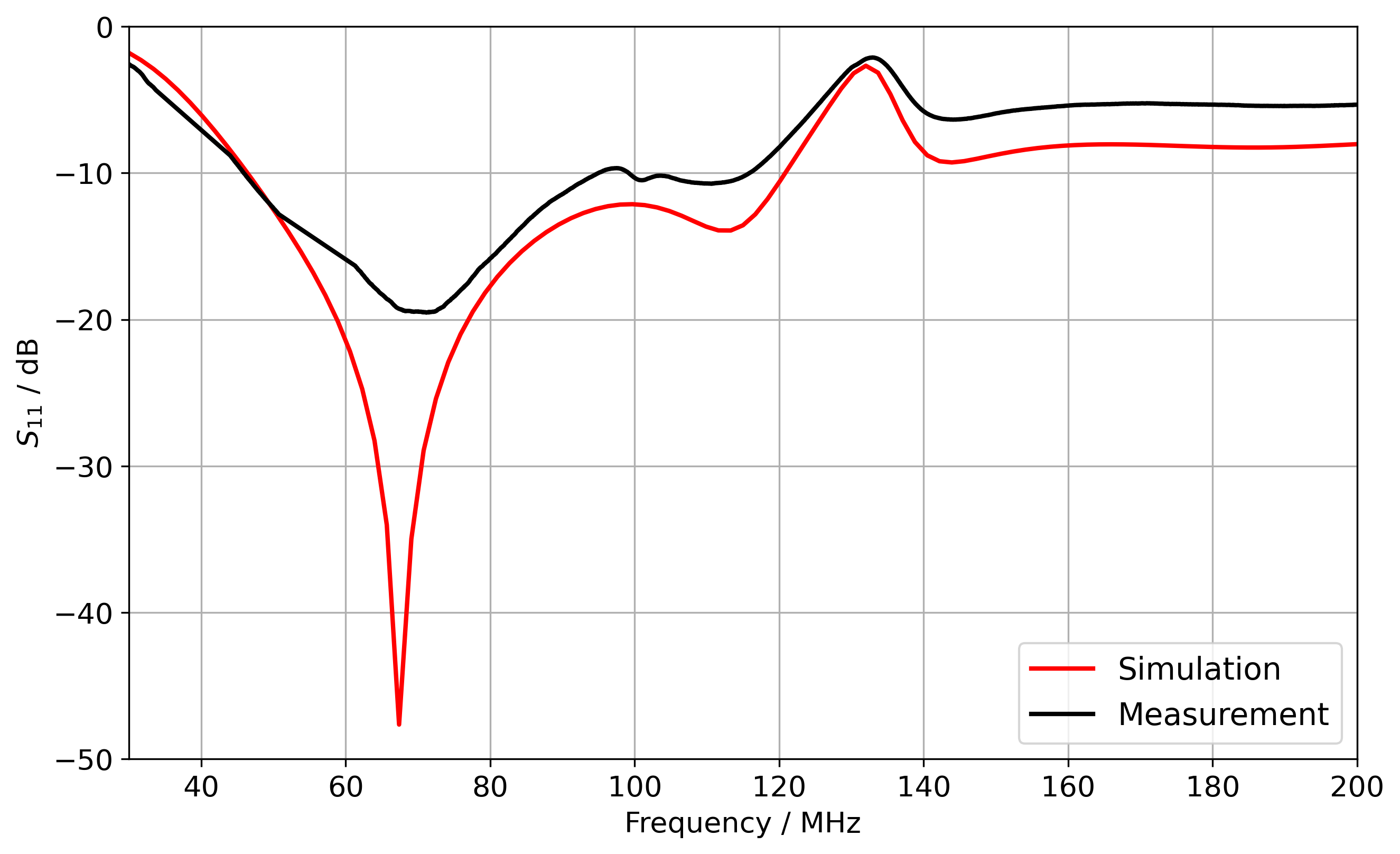}
}
\quad
\subfigure[Receiver]{
\includegraphics[width=0.5\columnwidth]{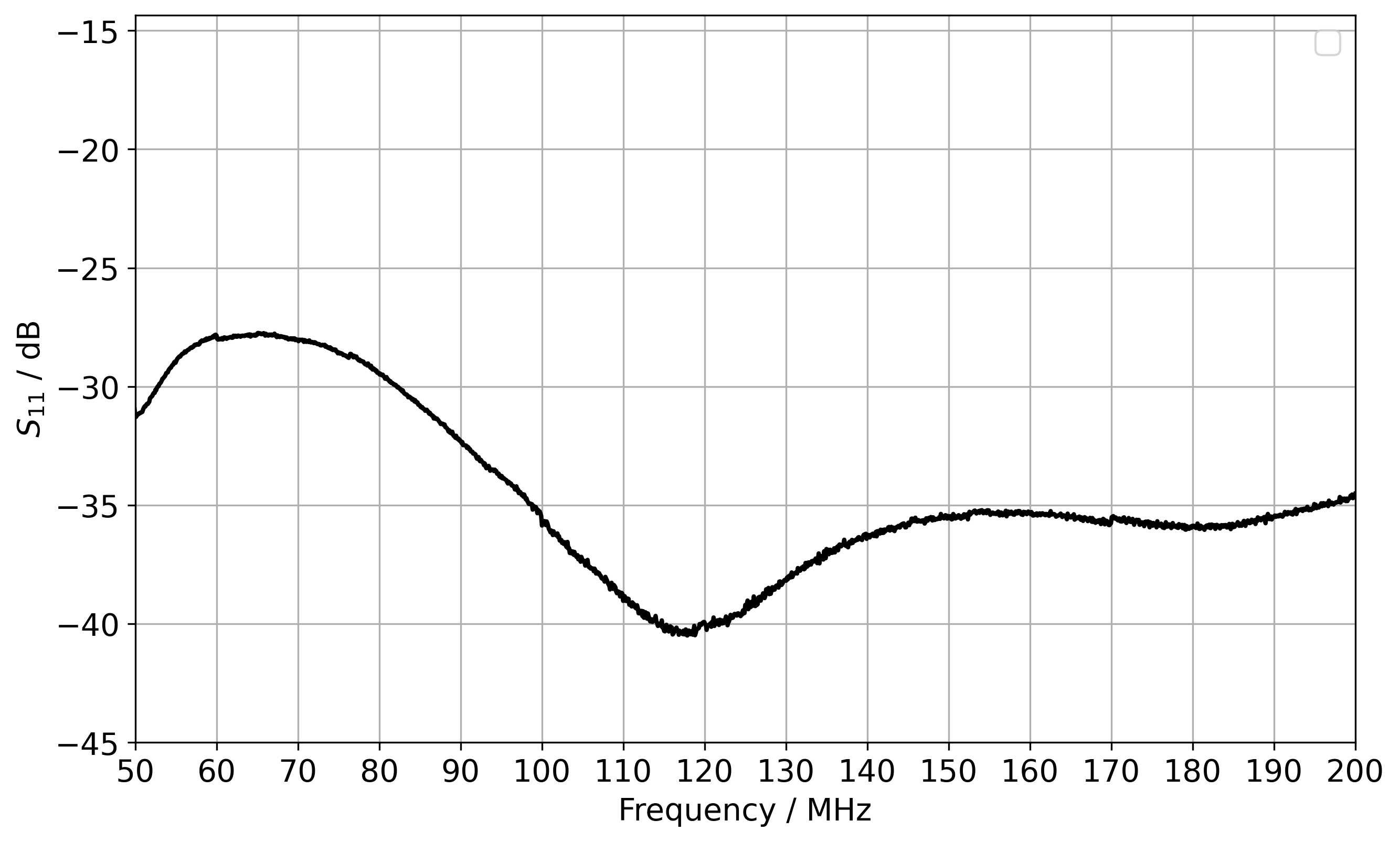}
}
\caption{In situ antenna and receiver measurement result.}
\label{fig:in situ measurement}
\end{figure}

After installation, we powered up the system. The $S_{11}$ coefficient of the antenna and receiver measured by its internal VNA are shown in Fig.~\ref{fig:in situ measurement}. For the antenna, the red line represents the measured values, while the black line denotes the simulated results for the antenna positioned above an 80~m ice sheet. The simulation and measurement results are generally consistent, affirming the validity of the design. The results of the receiver measurement are also consistent with our expectation.

\begin{figure*}[htbp]
\centering
\includegraphics[width=\columnwidth]{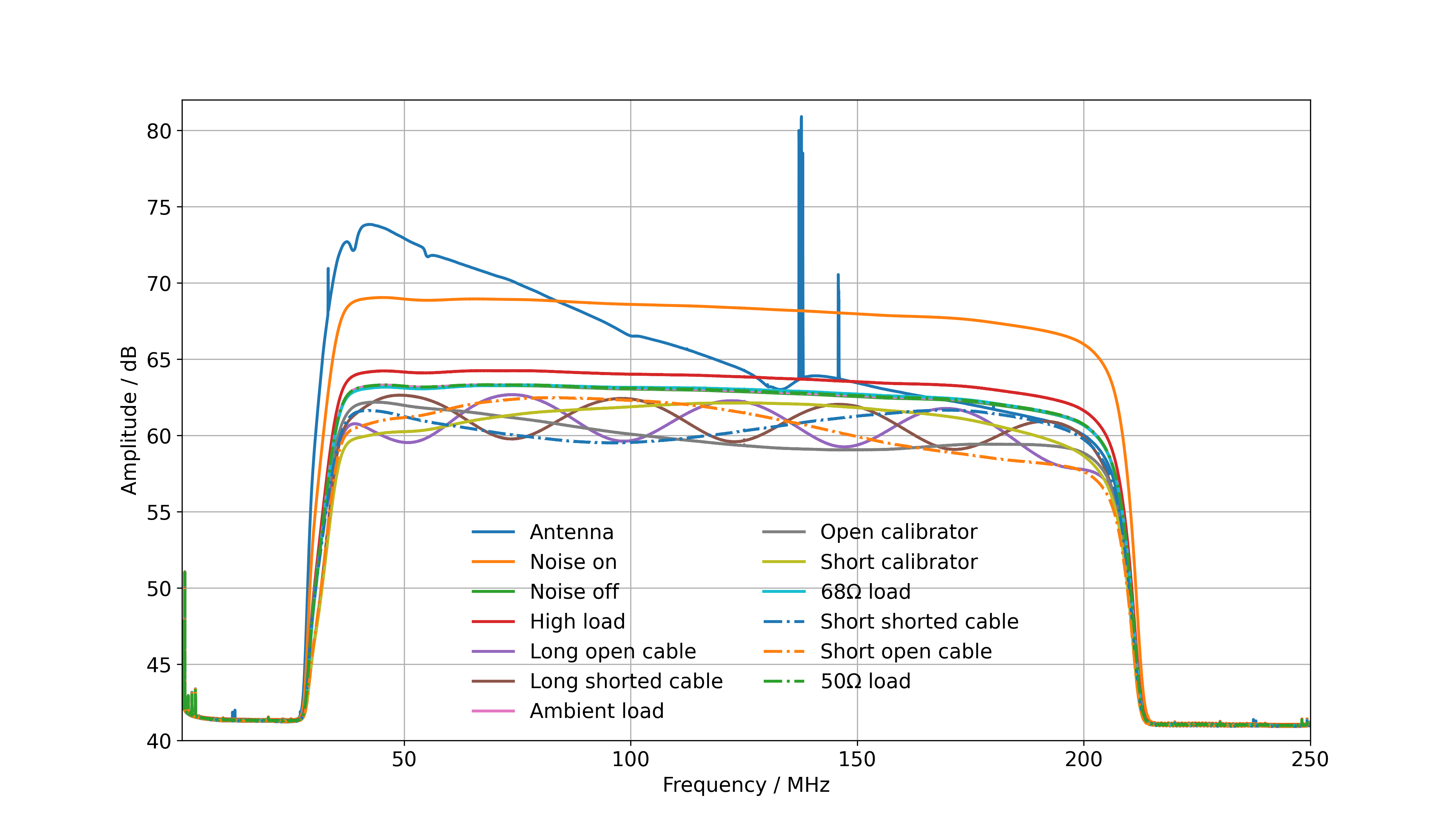}
\caption{The raw spectrums from the antenna and internal calibration spectrums. ``Antenna'' refers to the raw spectrums captured by antenna, ``Noise on" refers to the noise source activated, ``Noise off" refers to the noise source deactivated, ``High load" refers to the high-temperature load,  and ``Ambient load" refers to the ambient-temperature load.}
\label{fig:in situ calibration}
\end{figure*}

The calibrator source measurement results are shown in Fig.~\ref{fig:in situ calibration}, where the colored lines correspond to different following conditions. ``Antenna'' is the raw sky signal as observed. ``Noise on" and ``Noise off" are noise source activated and deactivated, used for relative calibration of the receiver. ``High load" and ``Ambient load" are spectra from high-temperature load and ambient-temperature load, used for absolute calibration of the receiver. ``Open calibrator", ``Short calibrator", ``Long open cable", ``Long shorted cable", ``Short shorted cable", and ``Short open cable" are calibrator spectrums for calculation of noise wave parameters\citep{price2023newtechniquemeasurenoise, sun2024calibration}. ``68$\Omega$ load" and ``50$\Omega$ load", used as antenna simulators to test instrument response. The strong interference captured by the antenna at 137~MHz and 145~MHz come from satellite radio transmissions. The shape and amplitude of these spectrums are similar to what we observe in the global spectrum experiments at our Xinjiang site~\citep{wu2024}, showing that the system is working normally. 

\section{Conclusions} 
\label{sec:conclusions}

The low level of RFI, relatively stable sky signal, and simple ground effects make the continental interior of Antarctica a promising site for precise low-frequency radio astronomy. We have carried out RFI and GPR surveys along the route from Zhongshan Station to Kunlun Station, and the results confirm our general expectations, showing that Antarctic sites do indeed have great potential for low-frequency radio astronomy in general, and for the 21~cm global spectrum measurement in particular. 

We have designed a set of instruments for the 21~cm global spectrum experiment (which comprises an elliptical dipole antenna with a receiver system). These are intended to operate in the  environmental conditions of Antarctica while ensuring high sensitivity and stability in detecting the faint 21~cm signals from the early universe. The self-calibration mechanisms indicate that the instrument has a sensitivity of 0.04~K in the expected frequency band. The system has been successfully deployed in Antarctica. Preliminary tests show resilience to the harsh environment, with a good alignment with simulated performance, confirming its capability to mitigate a number of frequently encountered challenges in such experiments, such as the ground effects. When the data from the 21~cm global spectrum experiment is returned (following the next annual expedition) and analyzed, it will further deepen our understanding of the characteristics of the Antarctic environment, and hopefully confirm that the Antarctica can provide an ideal site for low-frequency radio astronomy. Further data collection, longer observations, and refinements in instrumentation are essential to achieve a higher level of detection accuracy for faint cosmological signals. The results can also be compared with other sites, such as measurements taken in Xinjiang. This multi-site approach may help to mitigate site-specific systematic errors and offer a more comprehensive understanding of the experiments. 

\begin{acknowledgments}

We are grateful to the 40th and 41st Chinese National Antarctic Research Expedition team, supported by the Polar Research Institute of China and the Chinese Arctic and Antarctic Administration. This work is supported by the Chinese Academy of Science Key Instrument Grant ZDKYYQ20200008, and the National Natural Science Foundation of China under grant numbers 12473094 and 12273070. 
\end{acknowledgments}

%

\vspace{5mm}





\bibliography{bibliography}{}
\bibliographystyle{aasjournal}



\end{document}